\documentclass[10pt,a4paper]{article}

\usepackage{fullpage}

\usepackage[font=small,labelfont=small]{caption}

\newcommand\CC{C\nolinebreak[4]\hspace{-.05em}\raisebox{.4ex}{\relsize{-3}{\textbf{++}}}}

\usepackage{amsmath,amssymb,amsfonts,amsthm}
\usepackage{stmaryrd}
\usepackage{xfrac}
\usepackage{multirow}
\usepackage{colortbl}
\usepackage[table]{xcolor}
\usepackage{enumitem}
\usepackage{subcaption}

\usepackage{authblk}
\usepackage[numbers,sort&compress]{natbib}
\renewcommand{\cite}[1]{\citep{#1}}

\usepackage[ruled,vlined]{algorithm2e}
\SetAlFnt{\small}
\SetKwFor{ParFor}{parfor}{do}{endparfor}
\SetKwFor{EmptyLabel}{\:}{\:}{end}

\usepackage{color}
\definecolor{darkred}{rgb}{0.7,0.3,0.3}
\definecolor{darkblue}{rgb}{0.4,0.4,0.8}
\definecolor{darkgreen}{rgb}{0.1,0.5,0.1}
\definecolor{gray}{gray}{0.5}
\definecolor{lightgray}{gray}{0.75}
\definecolor{lightlightgray}{gray}{0.9}

\usepackage[pdftex=true,colorlinks=true,citecolor=darkred,linkcolor=darkblue]{hyperref}

\newcommand{\Hz}{~\textrm{Hz}}

\newcommand{\km}{~\textrm{km}}
\newcommand{\ms}{~\textrm{ms}}
\newcommand{\mus}{~\:\mu\textrm{s}}
\newcommand{\s}{~\textrm{s}}
\newcommand{\D}{\mathsf{D}}

\newcommand{\eg}{\textit{e.g.}~}
\newcommand{\ie}{\textit{i.e.}~}

\newcommand{\jump}[1]{\left\llbracket{#1}\right\rrbracket}

\newcommand{\mat}[1]{ \boldsymbol{\bf #1 } }
\renewcommand{\vec}[1]{ \boldsymbol{\bf #1 } }

\def\dvp#1#2{\displaystyle\frac{\partial{#1}}{\partial{#2}}}

\def\ddvp#1#2{\displaystyle\frac{\partial^2{#1}}{\partial{#2}}}

\newcommand{\fields}{\text{fields}}

\newcommand{\blkU}{\text{blkU}}
\newcommand{\lin}{\text{lin}}

\newcommand{\tri}{\text{tri}}
\newcommand{\tet}{\text{tet}}
\newcommand{\bnd}{\text{bnd}}

\title{\Large A GPU-accelerated nodal discontinuous Galerkin method with high-order absorbing boundary conditions and corner/edge compatibility}

\author[1,2]{A. Modave\footnote{Corresponding author: axel.modave@ensta-paristech.fr}}
\author[3]{A. Atle}
\author[1,4]{J. Chan}
\author[1]{T. Warburton}
\affil[1]{Virginia Tech, Blacksburg, VA, USA}
\affil[2]{POEMS (UMR 7231 CNRS-ENSTA-INRIA), ENSTA Paristech, Palaiseau, France}
\affil[3]{TOTAL E\&P, Houston, TX, USA}
\affil[4]{Rice University, Houston, TX, USA}

\date{}

\begin{document}

\maketitle

\vspace{-0.5cm}

\begin{abstract}
Discontinuous Galerkin finite element schemes exhibit attractive features for accurate large-scale wave-propagation simulations on modern parallel architectures.
For many applications, these schemes must be coupled with non-reflective boundary treatments to limit the size of the computational domain without losing accuracy or computational efficiency, which remains a challenging task.
In this paper, we present a combination of a nodal discontinuous Galerkin method with high-order absorbing boundary conditions (HABCs) for cuboidal computational domains.
Compatibility conditions are derived for HABCs intersecting at the edges and the corners of a cuboidal domain.
We propose a GPU implementation of the computational procedure, which results in a multidimensional solver with equations to be solved on 0D, 1D, 2D and 3D spatial regions.
Numerical results demonstrate both the accuracy and the computational efficiency of our approach.
\end{abstract}


\setlength{\parskip}{4pt}


\section{Introduction}
\label{sec:intro}

Numerical simulation tools play an important role for solving a wide range of large-scale wave-like problems in fields as diverse as underwater acoustics, electromagnetic scattering and seismic imaging.
In this context, computational procedures based on discontinuous Galerkin finite element methods are very attractive.
These methods can provide accurate solutions to realistic transient wave-like problems thanks to heterogeneous, non-conforming and curvilinear meshes, high-order discontinuous basis functions, and stable formulations for complicated physical models (see \eg \cite{cockburn2000development,warburton2000application,hesthaven2007nodal,leger2014parallel,li2014analysis,mercerat2015nodal,schmitt2016dgtd,toulorge2010curved,warburton2013low,ye2016discontinuous}).
In addition, the discrete structure of the numerical schemes is well suited for efficient massively parallel computing on distributed memory architectures and modern many-core accelerators \cite{chan2016gpu,klockner2009nodal,godel2010scalability,modave2015nodal,modave2016gpu,seny2014efficient}.

A critical issue for the simulation of wave phenomena is to correctly account for radiation of waves at artificial boundaries of the computational domain.
Non-reflective boundary treatments must be incorporated into the discontinuous Galerkin formulations in order to simulate the outward propagation of signals and perturbations generated from within the computational domain, even if they are not \textit{a priori} known.
The challenge then consists of devising boundary treatments that preserve the accuracy of the numerical solution without overpenalizing the computational efficiency of the implementation.

Basic boundary techniques encompass characteristic-based conditions, impedance conditions and sponge layers.
These techniques are robust, straightforward to implement and cheap to use, but they provide a relatively poor approximation of the solution.
Two families of techniques provide high-fidelity solutions at reasonable computational cost: perfectly matched layers (PMLs) (see \eg \cite{appelo2006perfectly,berenger1994perfectly,gedney1996anisotropic,hu2008development,komatitsch2003perfectly,lu2004discontinuous,modave2016perfectly}) and local high-order absorbing boundary conditions (HABCs) (see \eg \cite{baffet2012long,becache2010high,givoli2001high,givoli2004high,guddati2000continued,hagstrom1999radiation,hagstrom2004new,hagstrom2009complete,rabinovich2011finite}).
In the last two decades, PMLs have clearly received much more attention than HABCs.
One reason is that the PMLs are easier to implement than HABCs.
While HABCs require specific resolution procedures, with often cumbersome treatments of the corners of the computational domain, the PMLs can be rather easily implemented in existing computational codes with straightforward treatments at the corners.
Nevertheless, the accuracy of PMLs strongly depends on both the discretization and a selection of parameters.
Although procedures have been proposed to automate the selection \cite{collino1998optimizing,bermudez2007optimal,modave2014optimizing}, the parameters are often chosen by experimentation, which does not ensure an optimum accuracy.
By contrast, the parameters of HABCs can be tuned without any experiment thanks to reflection coefficients \cite{engquist1977absorbing,engquist1979radiation,hagstrom2004new,higdon1986absorbing} and \textit{a priori} error estimates \cite{hagstrom2009complete,hagstrom2010radiation} that allow for error analyses which are relevant for discretized problems.

With the aim at devising solvers that are both accurate and computationally efficient, we are interested in the coupling of discontinuous Galerkin methods with HABCs.
Early HABCs have been written with high-order partial derivatives in time and space \cite{engquist1977absorbing,engquist1979radiation,higdon1986absorbing}, where the order of derivatives is as high as the order of approximation.
Such formulations have been successfully implemented with finite element methods for time-harmonic problems  \cite{givoli1997high,schmidt2015non}, but their applicability is limited to low orders because of the high-order derivatives that must be discretized.
As an alternative strategy, Collino \cite{collino1993high,collino1993conditions} proposed a formulation with only low-order derivatives and auxiliary fields defined on the boundary, enabling the use of HABCs with high orders.
Hagstrom and Warburton \cite{hagstrom2004new,hagstrom2009complete} have incorporated such formulations in nodal discontinuous Galerkin schemes for time-dependent problems by rewriting the auxiliary fields with characteristic variables.
In order to deal with HABCs intersecting at the corners of computational domains, they also proposed a corner treatment based on compatibility conditions that preserve the accuracy of the solution.
However, the resulting solvers are rather complicated to generalize and to implement.
To the best of our knowledge, they have never been applied in 3D.
This has motivated the introduction of the double absorbing boundary (DAB) \cite{hagstrom2014double,baffet2014double} which simplifies the treatment of corners in HABC procedures.
LaGrone and Hagstrom \cite{lagrone2016double} recently proposed a 3D finite difference scheme with compatibility treatments for the edges and the corners of cuboidal domains.
The DAB technique relies on an extension of the domain with a thin layer where the auxiliary fields of the HABCs are defined.
In contrast with the strategy used in \cite{hagstrom2004new,hagstrom2009complete}, where linear systems are to be solved on the edges and the corners, the DAB leads to a purely iterative computational procedure.
Nevertheless, a larger number of discrete unknowns is required since the auxiliary fields are solved in a layer instead of only on the boundary.

In this paper, we propose a HABC procedure coupled with a nodal discontinuous Galerkin method for efficient 3D acoustic wave simulations in cuboidal domains.
Following the early works of Hagstrom and Warburton, HABCs are prescribed on the faces of the domain and compatibility conditions are derived for the edges and the corners, but we consider a specific HABC representation that leads to compatibility conditions which are easier to use.
While the auxiliary fields are governed by recursive equations in the representation considered in \cite{hagstrom2004new,hagstrom2009complete}, we use a representation close to the one proposed by Collino \cite{collino1993high,collino1993conditions} with uncoupled equations.
In addition to a simplification of the procedure, the obtained compatibility conditions overcome some inconsistencies that appear with previous formulations when deriving discontinuous Galerkin schemes.

In order to demonstrate the computational efficiency of our approach, we describe the implementation of the HABC procedure in a state-of-the-art GPU-accelerated discontinuous Galerkin solver and present results of a 3D realistic benchmark.
The complete procedure requires a multidimensional solver with equations posed in the volume, on the faces, the edges and the corners of the domain.
We use elaborate implementation techniques in order to improve the computational performance, while keeping the compatibility of the final implementation with implementations in the literature \cite{chan2016gpu,modave2015nodal}.

This paper is organized as follows.
In section \ref{sec:HABCform}, the HABCs are presented and edge/corner compatibility conditions are derived for both the wave equation and the pressure-velocity system.
Section \ref{sec:scheme} is dedicated to numerical schemes and implementation strategies.
We describe the discontinuous Galerkin finite element scheme, the low-storage Runge-Kutta scheme, and key aspects of the GPU-accelerated implementation.
In section \ref{sec:results}, we present 3D numerical results which validate the formulation and demonstrate the computational performance and the applicability of the approach.


\section{Non-reflective boundary treatment for cuboidal domain}
\label{sec:HABCform}

In this section, we derive high-order absorbing boundary conditions (HABCs) with compatibility conditions for edges and corners to simulate the propagation of waves in the infinite space with a cuboidal computational domain.
We aim at coupling these conditions with a numerical scheme based on the pressure-velocity system
\begin{subequations}
\begin{align}
  \dvp{p}{t} + \rho c^2 \nabla\cdot\vec{u} &= 0, \label{eqn:pressureEqn} \\
  \rho \dvp{\vec{u}}{t} + \nabla p &= 0, \label{eqn:velocityEqn}
\end{align}
\end{subequations}
where $p(t,\vec{x})$ is the pressure field, $\vec{u}(t,\vec{x})$ is the velocity field, $\rho$ is the density and $c$ is the phase velocity.
In this section, $\rho$ and $c$ are assumed to be constant.
When deriving the conditions, it is however more convenient to work with the wave equation
\begin{align}
  \ddvp{p}{t^2} - c^2 \Delta p = 0. \label{eqn:waveEqn}
\end{align}
The pressure-velocity system can be recovered by introducing the velocity field governed by equation \eqref{eqn:velocityEqn} and integrating the wave equation.

We first derive HABCs for a semi-infinite domain with a planar boundary (section \ref{sec:form:HABC}).
When using the obtained HABCs on all the faces of a cuboidal domain, a special treatment must be applied to the edges and the corners.
A treatment based on accuracy-preserving compatibility conditions is proposed in section \ref{sec:form:corners}.
The obtained equations of both the HABCs and the compatibility conditions involve second-order partial derivatives.
In section \ref{sec:form:final}, we derive equivalent formulations written using only first-order partial derivatives, which can quite naturally be coupled with spatial schemes based on the pressure-velocity system.
Mixed boundary conditions are briefly discussed in section \ref{sec:form:mixed}.


\subsection{High-order absorbing boundary conditions for planar boundary}
\label{sec:form:HABC}

Let us consider the half-space problem defined on the domain $\Omega = \{\vec{x}\in\mathbb{R}^3: x<0\}$ with the planar boundary $\Gamma = \{\vec{x}\in\mathbb{R}^3: x=0\}$, where $x$ is the coordinate in the Cartesian direction $\vec{e}_x$.
We seek a non-reflective boundary condition to prescribe on $\Gamma$.
For convenience, the transverse component of the position is denoted $\vec{y}$, such that $\vec{x}=(x,\vec{y})$.

The exact non-reflective boundary condition of the half-space problem is well-known.
Using notations borrowed from the pseudo-differential theory, it reads (see \eg \cite{engquist1977absorbing,hagstrom1999radiation})
\begin{equation}
  \mathcal{B}^x p = 0, \label{eqn:exactBC}
\end{equation}
with the pseudo-differential operator
\begin{align}
  \mathcal{B}^x &\equiv c\:\partial_x + \partial_t \sqrt{1 - c^2\Delta_\perp^x/\partial_{tt}}, \label{eqn:exactBCoperator}
\end{align}
where $\Delta_\perp^x \equiv \Delta - \partial_{xx}$ is the Laplace-Beltrami operator defined in the plan tangent to the direction $\vec{e}_x$.
Unfortunately, this condition is non-local in both time and space because of the square root, which makes it an impractical boundary treatment.

Local absorbing boundary conditions are classically obtained by approximating the square root $\sqrt{1+X}$ to localize the operator \eqref{eqn:exactBCoperator}.
The features of the obtained conditions depend on the approximation that is used for the square root.
In their seminal work, Engquist and Majda \cite{engquist1977absorbing} showed that Pad\'e approximations lead to stable conditions, while polynomial approximations based on Taylor expansions can lead to unstable conditions.
Other rational approximations have been used to derive one-way wave equations and absorbing boundary conditions with better accuracy for grazing waves, evanescent modes or long-duration simulations (see \eg \cite{antoine2006improved,asvadurov2003optimal,hagstrom2009complete,hagstrom2010radiation,halpern1988wide,higdon1986absorbing,ingerman2000optimal,kechroud2005numerical,lu1998complex,milinazzo1997rational}).
In this work, we restrict ourselves to the Pad\'e approximation, which corresponds to an asymptotic case for these rational approximations.
The boundary treatment and the computational procedure described hereafter will be extended to other approximations in the future.

The $(2N+1)^\text{th}$-order Pad\'e approximation of the square root $\sqrt{1+X}$ is classically written as the rational function \cite{bamberger1988higher,guan1985high,lu1998complex}
\begin{align}
  f_N(X) &= 1 + \frac{2}{M} \sum_{n=1}^N \frac{a_n X}{1 + b_n X}, \label{eqn:approxRoot1}
\end{align}
which we rewrite as
\begin{align}
  f_N(X) &= 1 + \frac{2}{M} \sum_{n=1}^N c_n \left(1 - \frac{1+c_n}{1+c_n + X}\right), \label{eqn:approxRoot2}
\end{align}
where $a_n = \sin^2(n\pi/M)$, $b_n = \cos^2(n\pi/M)$, $c_n = \tan^2(n\pi/M)$ and $M=2N+1$.
Using the formula \eqref{eqn:approxRoot2} to approximate the square root in the exact boundary operator \eqref{eqn:exactBCoperator} gives to the approximate boundary condition
\begin{equation}
  \mathcal{L}^x p = 0, \label{eqn:approxBC}
\end{equation}
with the pseudo-differential operator
\begin{align*}
  \mathcal{L}^x &\equiv \partial_t + c \: \partial_x + \frac{2}{M} \sum_{n=1}^{N} c_n \: \partial_t \left(1 - \frac{(1 + c_n) \partial_{tt}}{(1 + c_n) \partial_{tt} - c^2 \Delta_\perp^x}\right).
\end{align*}
We introduce $N$ auxiliary fields $p_n$ defined on the boundary $\Gamma$ as
\begin{align*}
  p_n \equiv \mathcal{M}_n^x\:p, \quad\quad \text{for } n=1,\dots,N,
\end{align*}
with
\begin{align}
  \mathcal{M}_n^x &\equiv \frac{(1 + c_n) \partial_{tt}}{(1 + c_n) \partial_{tt} - c^2 \Delta_\perp^x}, \quad\quad \text{for } n=1,\dots,N.
  \label{eqn:operatorM}
\end{align}
We can then explicitly rewrite the boundary condition \eqref{eqn:approxBC} as
\begin{align}
  \partial_t p + c \: \partial_x p &= \frac{2}{M} \sum_{n=1}^{N} c_n \partial_t (p_n-p), \label{eqn:HABCplan1}
\end{align}
where the auxiliary fields are governed by
\begin{align}
  \left(1+c_n\right) \partial_{tt} \left(p_n - p\right) - c^2 \Delta_\perp^x p_n &= 0, \quad\quad \text{for } n=1,\dots,N. \label{eqn:HABCplan2}
\end{align}
The boundary condition is local and requires the computation of the $N$ auxiliary equations on the boundary $\Gamma$.
Increasing the order of the Pad\'e approximation increases the number of auxiliary equations and the computational cost, but it also improves the accuracy for outgoing traveling waves \cite{collino1993conditions,collino1993high,engquist1977absorbing}.

We note that rational approximations of the square root can be written in several ways, leading to different representations of the boundary conditions.
Collino \cite{collino1993conditions,collino1993high} used the rational representation \eqref{eqn:approxRoot1} and obtained HABCs with auxiliary equations very close to equations \eqref{eqn:HABCplan1}-\eqref{eqn:HABCplan2}.
This representation can also be used with different parameters as soon as they verify some relations \cite{ha1994stability}.
Alternatively, rational functions written as continued fractions lead to HABCs with auxiliary fields governed by coupled equations (see \eg \cite{guddati2000continued,hagstrom2004new,asvadurov2003optimal,givoli2003high}).
In this work, we choose the representation \eqref{eqn:approxRoot2} because it leads to compatibility conditions that are easier to incorporate in a discontinuous Galerkin framework than those obtained in previous works.
This aspect is discussed later in the text.

\subsection{Compatibility conditions at edges and corners}
\label{sec:form:corners}

\begin{figure}[!t]
\centering
\begin{tabular}{c@{\:}c}
\begin{subfigure}[b]{6.5cm}
  \centering
  \caption{HABC defined on planes}
  \includegraphics[width=65mm]{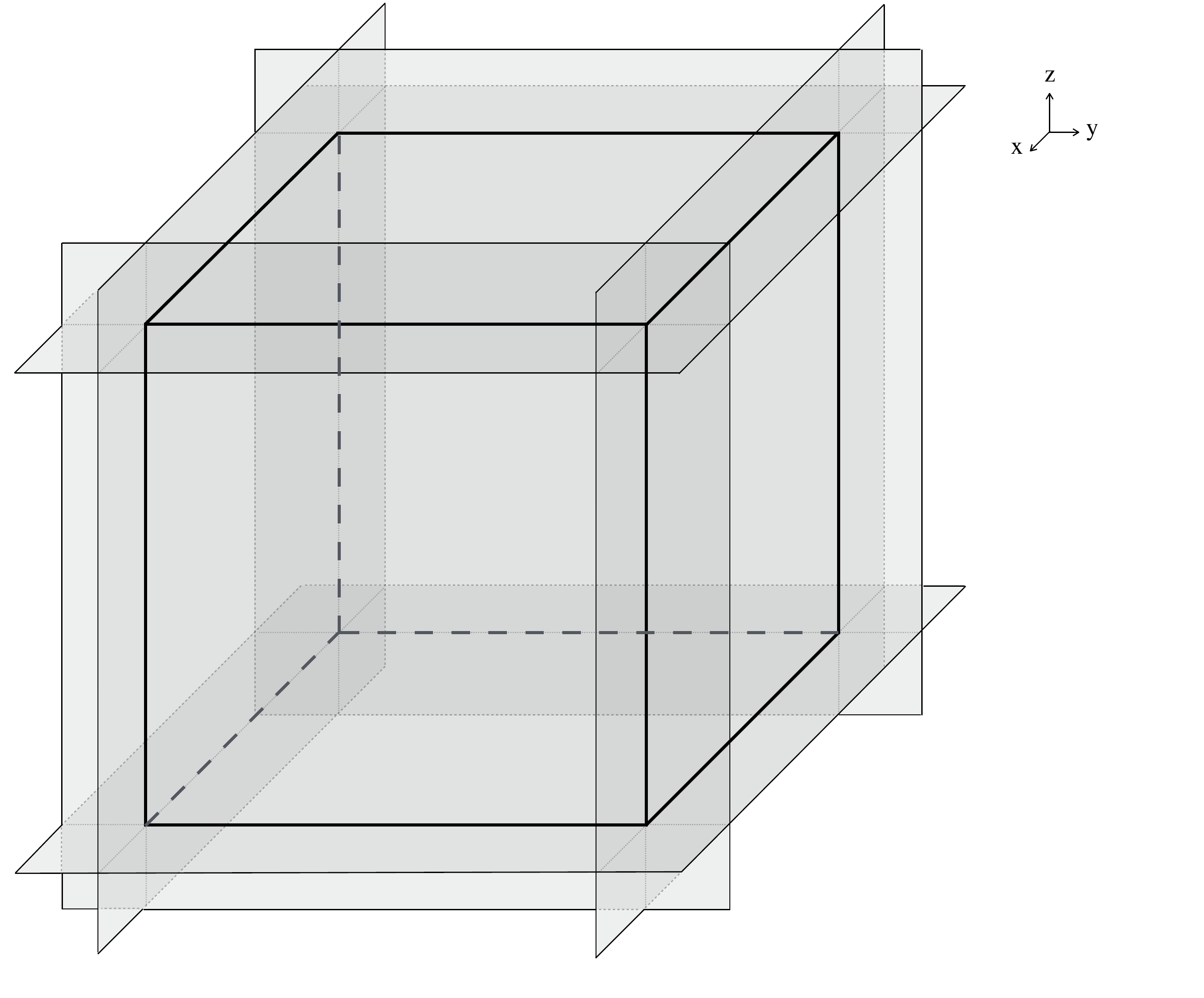}
  \label{fig:illustPlanes:1}
\end{subfigure}
&
\begin{subfigure}[b]{6.5cm}
  \centering
  \caption{HABC defined on faces}
  \includegraphics[width=65mm]{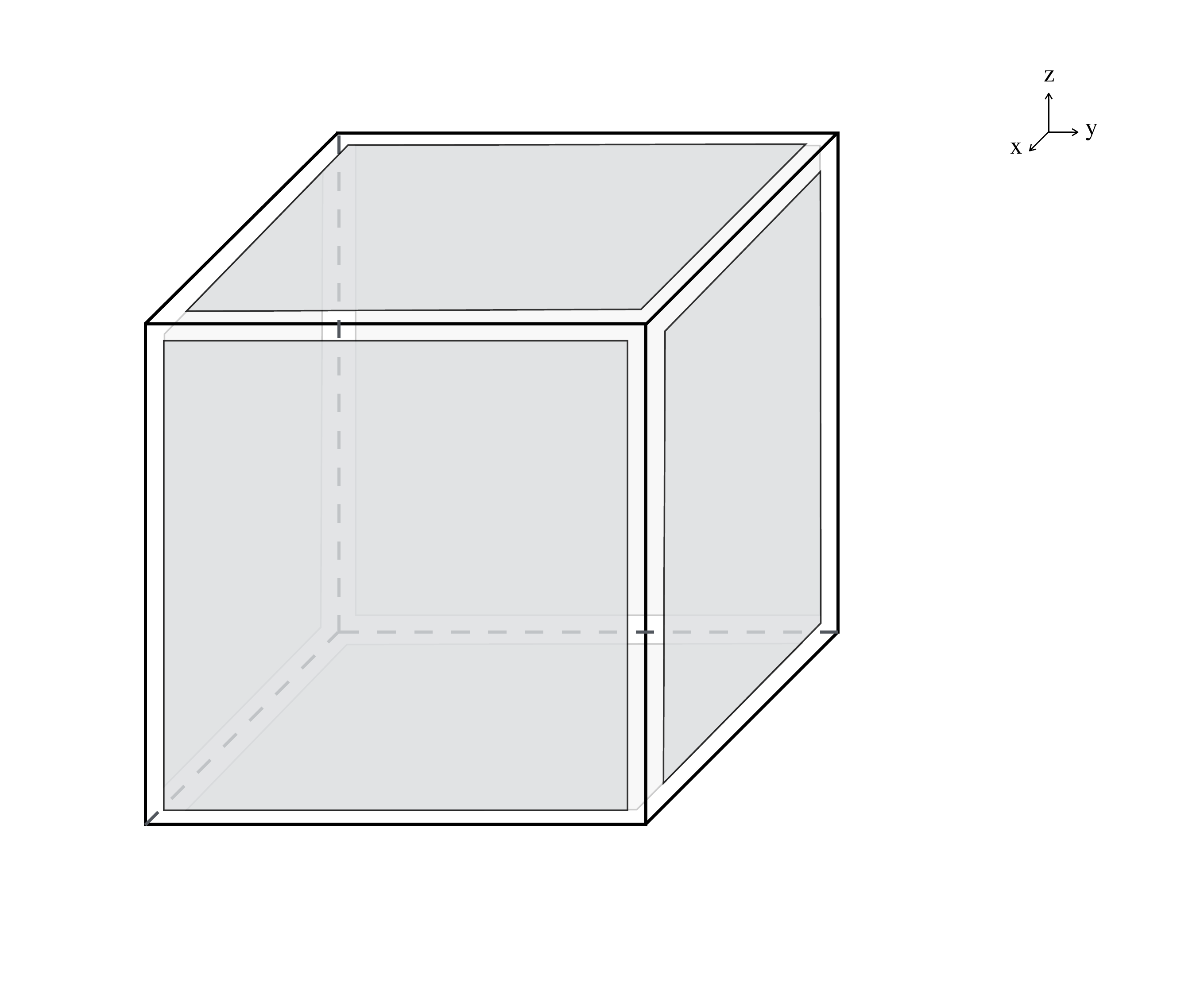}
  \label{fig:illustPlanes:2}
\end{subfigure}
\end{tabular}
\vspace{-0.5cm}
\caption{Illustration of HABCs defined on the planes containing the faces of the cuboidal domain (a) or defined only on the faces (b). When defining HABCs only on the faces, supplementary boundary conditions must be prescribed for the auxiliary variables on the borders of each face.}
\label{fig:illustPlanes}
\end{figure}

We now extend the boundary treatment to the borders of a cuboidal domain $\Omega=[-L^x,L^x]\times[-L^y,L^y]\times[-L^z,L^z]$ to simulate the infinite space $\mathbb{R}^3$.
The initial conditions and any source are assumed to be compactly supported inside the domain $\Omega$.
Under this assumption, the exact boundary condition \eqref{eqn:exactBC} and its approximation \eqref{eqn:approxBC} can be prescribed on the planes containing each of the six faces of the domain (figure \ref{fig:illustPlanes:1}).
In practice, we would like to prescribe the boundary condition \eqref{eqn:HABCplan1} only on the faces (figure \ref{fig:illustPlanes:2}).
However, computing the auxiliary equations \eqref{eqn:HABCplan2} only on the faces requires boundary conditions for the auxiliary fields on the borders of each face, which are on the edges of the domain.

To derive such conditions, we have adapted a strategy proposed by Hagstrom and Warburton \cite{hagstrom2004new} and based on purely algebraic manipulations of the equations.
It gives edge compatibility conditions which preserve the accuracy of the solution, and require the computation of supplementary auxiliary fields governed by auxiliary equations on the edges.
Proceeding in a hierarchical fashion, these new equations require boundary conditions on the borders of each edge, which are at the corners of the domain.
Algebraic manipulations of these equations similarly provide compatibility conditions with supplementary auxiliary fields at the corners.

In this section, we derive compatibility conditions for the edges belonging to the lines $(x,y)=(L^x,L^y)$, $(x,z)=(L^x,L^z)$ and $(y,z)=(L^y,L^z)$ and for the corner $(x,y,z)=(L^x,L^y,L^z)$.
The conditions for the other edges and corners can be obtained straightforwardly by symmetry.


\subsubsection*{Auxiliary fields}

For the sake of clarity, we first define the fields as
\begin{align}
  p_{ijk} &\equiv \mathcal{M}_i^x \mathcal{M}_j^y  \mathcal{M}_k^z \: p,
  \quad\quad \text{for } i,j,k=0,\dots,N,
  \label{eqn:defAddFields}
\end{align}
where the operators $\mathcal{M}_i^x$, $\mathcal{M}_j^y$, $\mathcal{M}_k^z$ are defined using equation \eqref{eqn:operatorM} for $i,j,k>0$ and $\mathcal{M}_0^x=\mathcal{M}_0^y=\mathcal{M}_0^z=1$.
The fields with zero, one, two and three non-zero subscripts are computed on the volume, the faces, the edges and the corners, respectively.
The field $p_{000}$ corresponds to the pressure field $p$.
The fields $p_{i00}$, $p_{0j0}$, $p_{00k}$, with $i,j,k=1,\dots,N$, are defined on the faces $x=L^x$, $y=L^y$, $z=L^z$, respectively.
The fields $p_{ij0}$, $p_{i0k}$, $p_{0jk}$, with $i,j,k=1,\dots,N$, are defined on the edges $(x,y)=(L^x,L^y)$, $(x,z)=(L^x,L^z)$, $(y,z)=(L^y,L^z)$, respectively.
The fields $p_{ijk}$, with $i,j,k=1,\dots,N$, are defined on the corner $(x,y,z)=(L^x,L^y,L^z)$.
Therefore, there are $N$, $N^2$ and $N^3$ auxiliary fields per face, edge and corner, respectively.
Note that, since the initial conditions on $p$ are assumed to be compactly supported inside the domain, the initial conditions on the auxiliary fields are equal to $0$.

The auxiliary fields have two properties that are used when deriving the compatibility conditions.
First, all the auxiliary fields satisfy the wave equation,
\begin{equation}
  \partial_{tt} p_{ijk} - c^2 \Delta p_{ijk} = 0,
  \quad\quad \text{for } i,j,k=0,\dots,N.
  \label{eqn:waveForPijk}
\end{equation}
Indeed, since the pressure field is in the kernel of the wave operator $\left[\partial_{tt}-c^2\Delta\right]$, the auxiliary fields defined by equation \eqref{eqn:defAddFields} are also in this kernel.
Then, the auxiliary fields are related by the following relations
\begin{subequations}
\begin{align}
  p_{ijk} &= \mathcal{M}_i^x \: p_{0jk}, \quad\quad \text{for } i,j,k=0,\dots,N, \label{eqn:relPijk_x} \\
  p_{ijk} &= \mathcal{M}_j^y \: p_{i0k}, \quad\quad \text{for } i,j,k=0,\dots,N, \label{eqn:relPijk_y} \\
  p_{ijk} &= \mathcal{M}_k^z \: p_{ij0}, \quad\quad \text{for } i,j,k=0,\dots,N. \label{eqn:relPijk_z}
\end{align}
\end{subequations}
These relations are trivially obtained by using definition \eqref{eqn:defAddFields} and observing that the operators $\mathcal{M}_i^x$, $\mathcal{M}_j^y$, $\mathcal{M}_k^z$ commute.


\subsubsection*{Boundary conditions on faces and 2D relations}

We write the boundary conditions for the 3D field on the faces $x=L^x$, $y=L^y$ and $z=L^z$.
On these faces, the field $p_{000}$ satisfies HABCs corresponding to half spaces,
\begin{align*}
  \mathcal{L}^x p_{000} &= 0, \quad\quad \text{for } x=L^x, \\
  \mathcal{L}^y p_{000} &= 0, \quad\quad \text{for } y=L^y, \\
  \mathcal{L}^z p_{000} &= 0, \quad\quad \text{for } z=L^z,
\end{align*}
which can be rewritten
\begin{subequations}
\begin{align}
  \partial_t p_{000} + c \: \partial_x p_{000}
    &= \frac{2}{M} \sum_{i=1}^{N} c_i \partial_t (p_{i00}-p_{000}), \quad\quad \text{for } x=L^x, \label{eqn:bc1:x} \\
  \partial_t p_{000} + c \: \partial_y p_{000}
    &= \frac{2}{M} \sum_{j=1}^{N} c_j \partial_t (p_{0j0}-p_{000}), \quad\quad \text{for } y=L^y, \label{eqn:bc1:y} \\
  \partial_t p_{000} + c \: \partial_z p_{000}
    &= \frac{2}{M} \sum_{k=1}^{N} c_k \partial_t (p_{00k}-p_{000}), \quad\quad \text{for } z=L^z. \label{eqn:bc1:z}
\end{align}
\end{subequations}
Using the definition \eqref{eqn:defAddFields}, we have the 2D relations
\begin{subequations}
\begin{align}
  (1+c_i) \partial_{tt} (p_{i00} - p_{000}) - c^2 (\partial_{yy} + \partial_{zz}) p_{i00}
    &= 0, \quad\quad \text{for } x=L^x, \label{eqn:lev1:iX} \\
  (1+c_j) \partial_{tt} (p_{0j0} - p_{000}) - c^2 (\partial_{xx} + \partial_{zz}) p_{0j0}
    &= 0, \quad\quad \text{for } y=L^y, \label{eqn:lev1:jY} \\
  (1+c_k) \partial_{tt} (p_{00k} - p_{000}) - c^2 (\partial_{xx} + \partial_{yy}) p_{00k}
    &= 0, \quad\quad \text{for } z=L^z, \label{eqn:lev1:kZ}
\end{align}
\end{subequations}
for $i,j,k = 1,\dots,N$.


\subsubsection*{Boundary conditions on edges and 1D relations}

On the borders of each face (\ie on the edges of the domain), the 2D fields need boundary conditions.
We derive here the compatibility conditions for the 2D fields on the edges $(x,y)=(L^x,L^y)$, $(x,z)=(L^x,L^z)$ and $(y,z)=(L^y,L^z)$.

Because $p_{000}$ belongs to the kernel of $\mathcal{L}^x$ and $\mathcal{L}^y$ on the edge $(x,y)=(L^x,L^y)$ and because these operators commute with $\mathcal{M}^x$ and $\mathcal{M}^y$, the 2D fields $p_{0j0}$ and $p_{i00}$ also belong to the kernel of $\mathcal{L}^x$ and $\mathcal{L}^y$ seen their definition.
This result gives boundary conditions for the 2D fields on the edge $(x,y)=(L^x,L^y)$.
Using similar reasoning, we obtain boundary conditions on the other edges.
We then have
\begin{align*}
  \mathcal{L}^x p_{0j0} &= 0, \quad\quad \text{for } (x,y)=(L^x,L^y), \\
  \mathcal{L}^y p_{i00} &= 0, \quad\quad \text{for } (x,y)=(L^x,L^y), \\
  \mathcal{L}^x p_{00k} &= 0, \quad\quad \text{for } (x,z)=(L^x,L^z), \\
  \mathcal{L}^z p_{i00} &= 0, \quad\quad \text{for } (x,z)=(L^x,L^z), \\
  \mathcal{L}^y p_{00k} &= 0, \quad\quad \text{for } (y,x)=(L^y,L^z), \\
  \mathcal{L}^z p_{0j0} &= 0, \quad\quad \text{for } (y,x)=(L^y,L^z),
\end{align*}
which can be rewritten
\begin{subequations}
\begin{align}
  \partial_t p_{0j0} + c\: \partial_x p_{0j0} &= \frac{2}{M} \sum_{i=1}^{N} c_i \partial_t (p_{ij0}-p_{0j0}),
    \quad\quad \text{for } (x,y)=(L^x,L^y), \label{eqn:bc2:1} \\
  \partial_t p_{i00} + c\: \partial_y p_{i00} &= \frac{2}{M} \sum_{j=1}^{N} c_j \partial_t (p_{ij0}-p_{i00}),
    \quad\quad \text{for } (x,y)=(L^x,L^y), \label{eqn:bc2:2} \\
  \partial_t p_{00k} + c\: \partial_x p_{00k} &= \frac{2}{M} \sum_{i=1}^{N} c_i \partial_t (p_{i0k}-p_{00k}),
    \quad\quad \text{for } (x,z)=(L^x,L^z), \label{eqn:bc2:3} \\
  \partial_t p_{i00} + c\: \partial_z p_{i00} &= \frac{2}{M} \sum_{k=1}^{N} c_k \partial_t (p_{i0k}-p_{i00}),
    \quad\quad \text{for } (x,z)=(L^x,L^z), \label{eqn:bc2:4} \\
  \partial_t p_{00k} + c\: \partial_y p_{00k} &= \frac{2}{M} \sum_{j=1}^{N} c_j \partial_t (p_{0jk}-p_{00k}),
    \quad\quad \text{for } (y,z)=(L^y,L^z), \label{eqn:bc2:5} \\
  \partial_t p_{0j0} + c\: \partial_z p_{0j0} &= \frac{2}{M} \sum_{k=1}^{N} c_k \partial_t (p_{0jk}-p_{0j0}),
    \quad\quad \text{for } (y,z)=(L^y,L^z), \label{eqn:bc2:6}
\end{align}
\end{subequations}
for $i,j,k = 1,\dots,N$.

We next derive governing equations for the 1D fields on the edges.
Using equations \eqref{eqn:relPijk_x}-\eqref{eqn:relPijk_z} to connect the 1D and the 2D fields, one has
\begin{subequations}
\begin{align}
  \partial_{tt} (1+c_i) (p_{ij0} - p_{0j0}) - c^2 (\partial_{yy} + \partial_{zz}) p_{ij0} &= 0,
    \quad\quad \text{for } (x,y)=(L^x,L^y), \label{eqn:lev2:ijX} \\
  \partial_{tt} (1+c_j) (p_{ij0} - p_{i00}) - c^2 (\partial_{xx} + \partial_{zz}) p_{ij0}&= 0,
    \quad\quad \text{for } (x,y)=(L^x,L^y), \label{eqn:lev2:ijY} \\
  \partial_{tt} (1+c_i) (p_{i0k} - p_{00k}) - c^2 (\partial_{yy} + \partial_{zz}) p_{i0k} &= 0,
    \quad\quad \text{for } (x,z)=(L^x,L^z), \label{eqn:lev2:ikX} \\
  \partial_{tt} (1+c_k) (p_{i0k} - p_{i00}) - c^2 (\partial_{xx} + \partial_{yy}) p_{i0k} &= 0,
    \quad\quad \text{for } (x,z)=(L^x,L^z), \label{eqn:lev2:ikZ} \\
  \partial_{tt} (1+c_j) (p_{0jk} - p_{00k}) - c^2 (\partial_{xx} + \partial_{zz}) p_{0jk} &= 0,
    \quad\quad \text{for } (y,z)=(L^y,L^z), \label{eqn:lev2:jkY} \\
  \partial_{tt} (1+c_k) (p_{0jk} - p_{0j0}) - c^2 (\partial_{xx} + \partial_{yy}) p_{0jk} &= 0,
    \quad\quad \text{for } (y,z)=(L^y,L^z), \label{eqn:lev2:jkZ}
\end{align}
\end{subequations}
for $i,j,k = 1,\dots,N$.
Unfortunately, these relations involve spatial derivatives that cannot be computed for 1D fields defined only on the edges.
For instance, only derivatives with respect to $z$ can be computed on the edge $(x,y)=(L^x,L^y)$, while equations \eqref{eqn:lev2:ijX} and \eqref{eqn:lev2:ijY} involve derivatives with respect to $y$ and $x$, respectively.
We then manipulate equations \eqref{eqn:lev2:ijX}-\eqref{eqn:lev2:jkZ} to eliminate such inadmissible derivatives.
Adding the equations corresponding to each edge and using equation \eqref{eqn:waveForPijk} give
\begin{subequations}
\begin{align}
  \partial_{tt} \left[(1+c_i+c_j) p_{ij0} - (1+c_i) p_{0j0} - (1+c_j) p_{i00}\right] - c^2 \partial_{zz} p_{ij0} &= 0,
    \quad\ \text{for } (x,y)=(L^x,L^y), \label{eqn:lev2:ijEdge} \\
  \partial_{tt} \left[(1+c_i+c_k) p_{i0k} - (1+c_i) p_{00k} - (1+c_k) p_{i00}\right] - c^2 \partial_{yy} p_{i0k} &= 0,
    \quad\ \text{for } (x,z)=(L^x,L^z), \label{eqn:lev2:ikEdge} \\
  \partial_{tt} \left[(1+c_j+c_k) p_{0jk} - (1+c_j) p_{00k} - (1+c_k) p_{0j0}\right] - c^2 \partial_{xx} p_{0jk} &= 0,
    \quad\ \text{for } (y,z)=(L^y,L^z), \label{eqn:lev2:jkEdge}
\end{align}
\end{subequations}
for $i,j,k = 1,\dots,N$.
These 1D relations involve only spatial derivatives which are well-defined on edges.


\subsubsection*{Boundary conditions at corners and 0D relations}

On the borders of each edge (\ie at the corners of the domain), the 1D field equations require boundary conditions.
We derive the compatibility conditions for the 1D fields at the corner $(x,y,z)=(L^x,L^y,L^z)$ by using a similar strategy than for the edges.

Because $p_{000}$ belongs to the kernel of $\mathcal{L}^x$, $\mathcal{L}^y$ and $\mathcal{L}^z$ and because these operators commute with $\mathcal{M}^x$, $\mathcal{M}^y$ and $\mathcal{M}^z$, the 1D fields $p_{0jk}$, $p_{i0k}$ and $p_{ij0}$ also belong to the kernel of $\mathcal{L}^x$, $\mathcal{L}^y$ and $\mathcal{L}^z$ by definition \eqref{eqn:defAddFields}.
At the corner $(x,y,z)=(L^x,L^y,L^z)$, we then have
\begin{align*}
  \mathcal{L}^x p_{0jk} &= 0, \\
  \mathcal{L}^y p_{i0k} &= 0, \\
  \mathcal{L}^z p_{ij0} &= 0,
\end{align*}
which can be rewritten
\begin{subequations}
\begin{align}
  \partial_t p_{0jk} + c \: \partial_x p_{0jk} &= \frac{2}{M} \sum_{i=1}^N c_i \partial_t (p_{ijk}-p_{0jk}),
  \label{eqn:bc3:1} \\
  \partial_t p_{i0k} + c \: \partial_y p_{i0k} &= \frac{2}{M} \sum_{j=1}^N c_j \partial_t (p_{ijk}-p_{i0k}),
  \label{eqn:bc3:2}  \\
  \partial_t p_{ij0} + c \: \partial_z p_{ij0} &= \frac{2}{M} \sum_{k=1}^N c_k \partial_t (p_{ijk}-p_{ij0}),
  \label{eqn:bc3:3} 
\end{align}
\end{subequations}
for $i,j,k = 1,\dots,N$.

We now derive relations for the 0D auxiliary fields at the corner.
Using equations \eqref{eqn:relPijk_x}-\eqref{eqn:relPijk_z} to connect the 0D and the 1D fields, one has
\begin{subequations}
\begin{align}
  (1+c_i) \partial_{tt} (p_{ijk} - p_{0jk}) - c^2 (\partial_{yy} + \partial_{zz}) p_{ijk} &= 0, \label{eqn:lev3:ijkX} \\
  (1+c_j) \partial_{tt} (p_{ijk} - p_{i0k}) - c^2 (\partial_{xx} + \partial_{zz}) p_{ijk} &= 0, \label{eqn:lev3:ijkY} \\
  (1+c_k) \partial_{tt} (p_{ijk} - p_{ij0}) - c^2 (\partial_{xx} + \partial_{yy}) p_{ijk} &= 0, \label{eqn:lev3:ijkZ}
\end{align}
\end{subequations}
for $i,j,k = 1,\dots,N$.
Unfortunately, these relations again involve spatial derivatives which are inadmissible if the 0D fields are defined only on the corner.
Again, we manipulate the equations to eliminate these derivatives.
Adding equations \eqref{eqn:lev3:ijkX}--\eqref{eqn:lev3:ijkZ}, removing the spatial derivatives by using equation \eqref{eqn:waveForPijk} and integrating in time give
\begin{align}
  (1+c_i+c_j+c_k) p_{ijk} - (1+c_i)p_{0jk} - (1+c_j)p_{i0k} - (1+c_k)p_{ij0} &= 0,
  \label{eqn:lev3:ijkCorner}
\end{align}
for $i,j,k = 1,\dots,N$.
These 0D relations do not involve any derivatives, and they can be used to remove the 0D fields $p_{ijk}$ from the boundary conditions \eqref{eqn:bc3:1}-\eqref{eqn:bc3:3}.

\subsection{Formulation with first-order partial differential equations}
\label{sec:form:final}

Since all the auxiliary fields $p_{ijk}(t,\vec{x})$ satisfy the wave equation \eqref{eqn:waveForPijk}, we can define auxiliary velocities $\vec{u}_{ijk}(t,\vec{x})$ such that each pair $(p_{ijk},\vec{u}_{ijk})$ satisfies the pressure-velocity system
\begin{subequations}
\begin{align}
  \partial_t p_{ijk} + \rho c^2 \: \nabla\cdot\vec{u}_{ijk} &= 0, \label{eqn:pressureEqnAux} \\
  \rho \: \partial_t \vec{u}_{ijk} + \nabla p_{ijk} &= 0,         \label{eqn:velocityEqnAux}
\end{align}
\end{subequations}
with $i,j,k=0,\dots,N$.
Introducing these auxiliary velocities and using integration in time on the equations derived in the previous section give a HABC formulation with only algebraic relations as boundary conditions and first-order differential equations as governing equations for the auxiliary fields:
\begin{itemize}

\item on the face $x=L^x$, the boundary condition \eqref{eqn:bc1:x} becomes
\begin{align}
  p_{000} - \rho c u_{000} &= \frac{2}{M} \sum_{i=1}^{N} c_i \: (p_{i00} - p_{000}),
  \label{eqn:HABCsyst:BClev1}
\end{align}
and the 2D fields are governed by
\begin{subequations}
\begin{align}
  (1+c_i) \partial_t p_{i00} + \rho c^2 \big( \partial_y v_{i00} + \partial_z w_{i00} \big) &= (1+c_i) \partial_t p_{000},
    \label{eqn:HABCsyst:GovLev1} \\
  \rho \partial_t v_{i00} + \partial_y p_{i00} &= 0, \label{eqn:HABCsyst:GovLev1b} \\
  \rho \partial_t w_{i00} + \partial_z p_{i00} &= 0, \label{eqn:HABCsyst:GovLev1c}
\end{align}
\end{subequations}
for $i=1,\dots,N$;

\item on the edge $(x,y)=(L^x,L^y)$, the boundary conditions \eqref{eqn:bc2:1}-\eqref{eqn:bc2:2} become
\begin{subequations}
\begin{align}
  p_{0j0} - \rho cu_{0j0} &= \frac{2}{M} \sum_{i=1}^{N} c_i (p_{ij0} - p_{0j0}),
  \label{eqn:HABCsyst:BClev2a} \\
  p_{i00} - \rho cv_{i00} &= \frac{2}{M} \sum_{j=1}^{N} c_j (p_{ij0} - p_{i00}),
  \label{eqn:HABCsyst:BClev2b}
\end{align}
\end{subequations}
and the 1D fields are governed by
\begin{subequations}
\begin{align}
  (1+c_i+c_j) \partial_t p_{ij0} + \rho c^2 \partial_z w_{ij0} &= (1+c_j) \partial_t p_{i00} + (1+c_i) \partial_t p_{0j0}
    \label{eqn:HABCsyst:GovLev2} \\
  \rho \partial_t w_{ij0} + \partial_z p_{ij0} &= 0,
    \label{eqn:HABCsyst:GovLev2b}
\end{align}
\end{subequations}
for $i,j=1,\dots,N$;

\item at the corner $(x,y,z)=(L^x,L^y,L^z)$, the boundary conditions \eqref{eqn:bc3:1}-\eqref{eqn:bc3:3} become
\begin{subequations}
\begin{align}
  p_{0jk} - \rho cu_{0jk} &= \frac{2}{M} \sum_{i=1}^{N} c_i \big( (1-C_{ijk}) p_{0jk} + C_{jki} p_{i0k} + C_{kij} p_{ij0} \big) ,
  \label{eqn:HABCsyst:BClev3a} \\
  p_{i0k} - \rho cv_{i0k} &= \frac{2}{M} \sum_{j=1}^{N} c_j \big( C_{ijk} p_{0jk} + (1-C_{jki}) p_{i0k} + C_{kij} p_{ij0} \big) ,
  \label{eqn:HABCsyst:BClev3b} \\
  p_{ij0} - \rho cv_{ij0} &= \frac{2}{M} \sum_{k=1}^{N} c_k \big( C_{ijk} p_{0jk} + C_{jki} p_{i0k} + (1-C_{kij}) p_{ij0} \big) ,
  \label{eqn:HABCsyst:BClev3c}
\end{align}
\end{subequations}
with $C_{ijk} = (1+c_i)/(1+c_i+c_j+c_k)$, for $i,j,k=1,\dots,N$.

\end{itemize}
Similar relations can be obtained for the other faces, edges and corners.

The boundary treatment presented in this section then consists of a multidimensional solver with equations to solve in the volume, on the faces, on the edges and at the corners of the domain.
The formulation is summarized in this way:
\begin{itemize}
\setlength\itemsep{0em}
\item on the volume, the 3D fields (pressure and velocity) are governed by the classical pressure-velocity system with initial conditions and/or sources which are compactly supported inside the domain;
\item on the faces, one boundary condition is prescribed using the algebraic relation \eqref{eqn:HABCsyst:BClev1} and $N$ sets of 2D auxiliary fields are governed by the first-order differential equations \eqref{eqn:HABCsyst:GovLev1}-\eqref{eqn:HABCsyst:GovLev1c};
\item on the edges, $2N$ boundary conditions ($N$ for each adjacent face) are prescribed using the algebraic relations \eqref{eqn:HABCsyst:BClev2a}-\eqref{eqn:HABCsyst:BClev2b} and $N^2$ sets of 1D auxiliary fields are governed by the first-order differential equations \eqref{eqn:HABCsyst:GovLev2}-\eqref{eqn:HABCsyst:GovLev2b};
\item at the corners, $3N^2$ boundary conditions ($N^2$ for each adjacent edge) are prescribed using the algebraic relations \eqref{eqn:HABCsyst:BClev3a}-\eqref{eqn:HABCsyst:BClev3c}.
\end{itemize}
The equation numbers correspond to the face $x=L^x$, the edge $(x,y)=(L^x,L^y)$ and the corner $(x,y,z)=(L^x,L^y,L^z)$.
By the assumption of the compact support of the initial condition inside the domain, the initial conditions for the auxiliary variable fields are all zero.


\subsection{Extension to mixed boundary conditions}
\label{sec:form:mixed}

We briefly address the case where a homogeneous boundary condition is prescribed on one or more faces of the computational domain.
In exploration geophysics, for instance, the computational cuboidal domain must represent the underground structure.
A HABC can be used on the lateral and bottom faces of the domain, while the so-called \textit{free-surface} boundary condition (which is $p=0$ in the acoustic model) must be prescribed on the upper face to simulate the Earth's surface.

A homogeneous boundary condition on the 3D fields is straightforwardly incorporated into the boundary procedure by using the same condition on the auxiliary 2D and 1D fields.
Indeed, let us consider the homogeneous Dirichlet condition $p=0$ on one face of the domain and a HABC on the adjacent faces.
Seen definition \eqref{eqn:defAddFields}, the 2D auxiliary fields of the adjacent faces must be set to zero on the edges if $p=0$.
Therefore, no auxiliary field must be computed on edges and corners belonging to faces where a homogeneous boundary condition is prescribed.


\section{Numerical scheme and computational implementation}
\label{sec:scheme}

In this section, we describe the explicit time-stepping procedure to solve the HABC formulation (section \ref{sec:scheme:Time}),
the numerical discretization with the discontinuous Galerkin time domain scheme (section \ref{sec:scheme:DG}) and the main components of our GPU-accelerated implementation (section \ref{sec:implementation}).

\subsection{Explicit time-stepping procedure}
\label{sec:scheme:Time}

Because of the coupling between the differential equations and the algebraic relations of the different levels, the formulation described in section \ref{sec:form:final} cannot be straightforwardly solved with an explicit time-stepping procedure.
For instance, the time derivatives of both 3D pressure and 2D pressures appear in equation \eqref{eqn:HABCsyst:GovLev1}, and the time derivatives of both 2D pressures and 1D pressures appear in equation \eqref{eqn:HABCsyst:GovLev2}.
In order to allow an explicit time-stepping procedure, we first reformulate the boundary conditions introducing temporary variables and using characteristic variables (as in \cite{hagstrom2004new}).
In the directions $\vec{e}_x$, $\vec{e}_y$, $\vec{e}_z$, the characteristic variables read
\begin{subequations}
\begin{align}
  r^{-x}_{ijk} &= p_{ijk} - \rho c \: u_{ijk}, &
  r^{+x}_{ijk} &= p_{ijk} + \rho c \: u_{ijk}, \label{eqn:proc:charact1} \\
  r^{-y}_{ijk} &= p_{ijk} - \rho c \: v_{ijk}, &
  r^{+y}_{ijk} &= p_{ijk} + \rho c \: v_{ijk}, \label{eqn:proc:charact2} \\
  r^{-z}_{ijk} &= p_{ijk} - \rho c \: w_{ijk}, &
  r^{+z}_{ijk} &= p_{ijk} + \rho c \: w_{ijk}, \label{eqn:proc:charact3} 
\end{align}
\end{subequations}
where $r^{-x}_{ijk}$, $r^{-y}_{ijk}$, $r^{-z}_{ijk}$ and $r^{+x}_{ijk}$, $r^{+y}_{ijk}$, $r^{+z}_{ijk}$ contain information traveling downwardly and upwardly, respectively, along the direction $\vec{e}_x$, $\vec{e}_y$, $\vec{e}_z$.
The comprehensive boundary formulation reads

\begin{itemize}

\item on the face $x=L^x$, temporary fields $p_{i00}^\star$ are defined as
\begin{align}
  p_{i00}^\star &= p_{i00} - (r^{-x}_{000}+r^{+x}_{000}),
  \label{eqn:proc:pi00Star}
\end{align}
the boundary condition gives
\begin{align}
  r^{-x}_{000} = \frac{1}{M} \sum_{i=1}^N c_i \: p_{i00}^\star,
  \label{eqn:proc:faceBC}
\end{align}
and the 2D relations becomes
\begin{subequations}
\begin{align}
  (1+c_i) \  \partial_t p_{i00}^\star + \rho c^2 \big( \partial_y v_{i00} + \partial_z w_{i00} \big) &= 0,
  \label{eqn:proc:2Dpvsyst1} \\
  \rho \partial_t v_{i00} + \partial_y p_{i00} &= 0,
  \label{eqn:proc:2Dpvsyst2} \\
  \rho \partial_t w_{i00} + \partial_z p_{i00} &= 0,
  \label{eqn:proc:2Dpvsyst3}
\end{align}
\end{subequations}
for $i=1,\dots,N$;

\item on the edge $(x,y)=(L^x,L^y)$, temporary fields $p_{ij0}^\star$ are defined as
\begin{align}
  p_{ij0}^\star &= p_{ij0} - \big( C_{ji}  (r^{-y}_{i00}+r^{+y}_{i00}) + C_{ij} (r^{-x}_{0j0}+r^{+x}_{0j0}) \big),
  \label{eqn:proc:pij0Star}
\end{align}
with $C_{ij} = (1+c_i)/(1+c_i+c_j)$, the boundary conditions give
\begin{subequations}
\begin{align}
  r^{-y}_{i00}
    &= \frac{1}{M} \sum_{j=1}^N c_j \: \big(p_{ij0}^\star
      + (C_{ji}-1) \: (r^{-y}_{i00}+r^{+y}_{i00})
      + C_{ij} \: (r^{-x}_{0j0}+r^{+x}_{0j0})\big),
  \label{eqn:proc:edgeBC1} \\
  r^{-x}_{0j0} 
    &= \frac{1}{M} \sum_{i=1}^N c_i \: \big(p_{ij0}^\star
      + C_{ji} \: (r^{-y}_{i00}+r^{+y}_{i00})
      + (C_{ij}-1) \: (r^{-x}_{0j0}+r^{+x}_{0j0})\big),
  \label{eqn:proc:edgeBC2}
\end{align}
\end{subequations}
and the 1D relations becomes
\begin{subequations}
\begin{align}
  (1+c_i+c_j) \ \partial_t p_{ij0}^\star + \rho c^2 \partial_z w_{ij0} &= 0,
  \label{eqn:proc:1Dpvsyst1} \\
  \rho \partial_t w_{ij0} + \partial_z p_{ij0} &= 0,
  \label{eqn:proc:1Dpvsyst2}
\end{align}
\end{subequations}
for $i,j=1,\dots,N$;

\item at the corner $(x,y,z)=(L^x,L^y,L^z)$, the boundary conditions give
\begin{subequations}
\begin{align}
  r^{-x}_{0jk}
  = \frac{1}{M} \sum_{i=1}^N c_i \: \big(
        (C_{ijk}-1) (r^{-x}_{0jk}+r^{+x}_{0jk})
      + C_{jki}     (r^{-y}_{i0k}+r^{+y}_{i0k})
      + C_{kij}     (r^{-z}_{ij0}+r^{+z}_{ij0}) \big),
  \label{eqn:proc:cornerBC1}
\\
  r^{-y}_{i0k}
  = \frac{1}{M} \sum_{j=1}^N c_j \: \big(
        C_{ijk}     (r^{-x}_{0jk}+r^{+x}_{0jk})
      + (C_{jki}-1) (r^{-y}_{i0k}+r^{+y}_{i0k})
      + C_{kij}     (r^{-z}_{ij0}+r^{+z}_{ij0}) \big),
  \label{eqn:proc:cornerBC2}
\\
  r^{-z}_{ij0}
  = \frac{1}{M} \sum_{k=1}^N c_k  \: \big(
        C_{ijk}     (r^{-x}_{0jk}+r^{+x}_{0jk})
      + C_{jki}     (r^{-y}_{i0k}+r^{+y}_{i0k})
      + (C_{kij}-1) (r^{-z}_{ij0}+r^{+z}_{ij0}) \big),
  \label{eqn:proc:cornerBC3}
\end{align}
\end{subequations}
for $i,j,k=1,\dots,N$.

\end{itemize}
Similar relations can be obtained for the other faces, edges and corners.

\begin{algorithm}[!t]
\caption{Explicit multidimensional solver for the pressure-velocity system with HABC and edge/corner compatibility.
The symbols and equation numbers correspond to face the $x=L^x$, the edge $(x,y)=(L^x,L^y)$ and the corner $(x,y,z)=(L^x,L^y,L^z)$.}
\label{algo:multiSolver}
\begin{minipage}{0.95\textwidth}
\medskip
\setlength\itemsep{0em}
\textit{3D solver (on the volume):}
\begin{itemize}
\setlength\itemsep{0em}
\item update the 3D fields $p_{000}$ and $\vec{u}_{000}$ at $t_2$ by solving the pressure-velocity system \eqref{eqn:pressureEqn}-\eqref{eqn:velocityEqn}, using the 3D incoming characteristics $r^{-x}_{000}$ computed at $t_1$ as boundary condition;
\end{itemize}

\textit{2D solver (on the faces):}
\begin{itemize}
\setlength\itemsep{0em}
\item compute the 2D temporary fields $p^\star_{i00}$ at $t_1$ by using equation \eqref{eqn:proc:pi00Star};
\item update the 2D fields $p^\star_{i00}$, $u_{i00}$ and $v_{i00}$ at $t_2$ by solving the 2D relations \eqref{eqn:proc:2Dpvsyst1}-\eqref{eqn:proc:2Dpvsyst3}, using the 2D incoming characteristics computed $r^{-y}_{i00}$ at $t_1$ as boundary condition;
\item compute the 3D outgoing characteristics $r^{+x}_{000}$ at $t_2$ by using equations \eqref{eqn:proc:charact1};
\item compute the 3D incoming characteristics $r^{-x}_{000}$ at $t_2$ by using equation \eqref{eqn:proc:faceBC};
\item compute the 2D pressure fields $p_{i00}$ at $t_2$ by reusing equation \eqref{eqn:proc:pi00Star};
\end{itemize}

\textit{1D solver (on the edges):}
\begin{itemize}
\setlength\itemsep{0em}
\item compute the 1D temporary fields $p^\star_{ij0}$ at $t_1$ by using equation \eqref{eqn:proc:pij0Star};
\item update the 1D fields $p^\star_{ij0}$ and $u_{ij0}$ at $t_2$ by solving the 1D relations \eqref{eqn:proc:1Dpvsyst1}-\eqref{eqn:proc:1Dpvsyst2}, using the 1D incoming characteristics $r^{-z}_{ij0}$ computed at $t_1$ as boundary condition;
\item compute the 2D outgoing characteristics $r^{+y}_{i00}$ and $r^{+x}_{0j0}$ at $t_2$ by using equations \eqref{eqn:proc:charact1}-\eqref{eqn:proc:charact2};
\item compute the 2D incoming characteristics $r^{-y}_{i00}$ and $r^{-x}_{0j0}$ at $t_2$ by solving the $2N$-equations system \eqref{eqn:proc:edgeBC1}-\eqref{eqn:proc:edgeBC2};
\item compute the 1D pressure fields $p_{ij0}$ at $t_2$ by reusing equation \eqref{eqn:proc:pij0Star};
\end{itemize}

\textit{0D solver (at the corners):}
\begin{itemize}
\setlength\itemsep{0em}
\item compute the 1D outgoing characteristics $r^{+x}_{0jk}$, $r^{+y}_{i0k}$ and $r^{+z}_{ij0}$ at $t_2$ by using \eqref{eqn:proc:charact1}-\eqref{eqn:proc:charact3};
\item  compute the 1D incoming characteristics $r^{-x}_{0jk}$, $r^{-y}_{i0k}$ and $r^{-z}_{ij0}$ at $t_2$ by solving the $3N^2$-equations system \eqref{eqn:proc:cornerBC1}-\eqref{eqn:proc:cornerBC3};
\end{itemize}
\medskip
\end{minipage}
\end{algorithm}

The numerical solution of this formulation can be computed with an explicit time-stepping scheme by solving the different levels successively at each time step, starting with the 3D and ending with the 0D.
The complete procedure to update the solution from time $t_1$ to time $t_2=t_1+\Delta t$ is sketched in algorithm \ref{algo:multiSolver}.
The computational load is mainly due to two kinds of operations:
\begin{enumerate}

\item
The 3D, 2D, 1D solvers solve first-order differential systems to update the corresponding pressure and velocity fields.
Since the systems of the faces (equations \eqref{eqn:proc:2Dpvsyst1}-\eqref{eqn:proc:2Dpvsyst3}) and the edges (equations \eqref{eqn:proc:1Dpvsyst1}-\eqref{eqn:proc:1Dpvsyst2}) resemble to pressure-velocity systems, we use the same numerical scheme to solve the systems over each dimension (\ie volume, faces and edges).
To update the 3D, 2D, 1D fields at $t=t_2$, the 3D, 2D, 1D solvers use boundary conditions based on incoming characteristics computed at $t=t_1$.
Such characteristic-based boundary conditions are naturally incorporated in discontinuous Galerkin formulations (see section \ref{sec:scheme:DG}).

\item
The 2D, 1D, 0D solvers update respectively the 3D, 2D, 1D incoming characteristics at $t=t_2$, using the fields which are already computed at $t=t_2$.
The computation is straightforward on the faces (equation \eqref{eqn:proc:faceBC}), but it requires the solution of linear systems with $2N$ unknowns on the edges (equations \eqref{eqn:proc:edgeBC1}-\eqref{eqn:proc:edgeBC2}) and $3N^2$ unknowns at the corners (equations \eqref{eqn:proc:cornerBC1}-\eqref{eqn:proc:cornerBC3}).
For instance, for the edge $(x,y)=(L^x,L^y)$, the $2N \times 2N$  system with equations \eqref{eqn:proc:edgeBC1}-\eqref{eqn:proc:edgeBC2} can be written
\begin{align}
  \left[\begin{array}{cc} M\vec{I}-\vec{C}^{(1)} & \vec{C}^{(2)} \\ \vec{C}^{(2)} & M\vec{I}-\vec{C}^{(1)} \end{array}\right]
  \left[\begin{array}{c} \vec{r}^{-x} \\ \vec{r}^{-y} \end{array}\right]
  =
  \left[\begin{array}{cc} \vec{C}^{(1)} & \vec{C}^{(2)} \\ \vec{C}^{(2)} & \vec{C}^{(1)} \end{array}\right]
  \left[\begin{array}{c} \vec{r}^{+x} \\ \vec{r}^{+y} \end{array}\right]
  +
  \left[\begin{array}{c} \vec{b}^{x} \\ \vec{b}^{y} \end{array}\right]
\label{eqn:systEdgeSymmetry}
\end{align}
with
\begin{align*}
  \vec{r}^{-x} &=
  \left[
    r^{-x}_{010} \;\;\;
    r^{-x}_{020} \;\;\;
    \cdots \;\;\;
    r^{-x}_{0N0}
  \right]^T,
&
  \vec{r}^{+x} &=
  \left[
    r^{+x}_{010} \;\;\;
    r^{+x}_{020} \;\;\;
    \cdots \;\;\;
    r^{+x}_{0N0}
  \right]^T,
\\
  \vec{r}^{-y} &=
  \left[
    r^{-y}_{100} \;\;\;
    r^{-y}_{200} \;\;\;
    \cdots \;\;\;
    r^{-y}_{N00}
  \right]^T,
&
  \vec{r}^{-y} &=
  \left[
    r^{+y}_{100} \;\;\;
    r^{+y}_{200} \;\;\;
    \cdots \;\;\;
    r^{+y}_{N00}
  \right]^T,
\end{align*}
\begin{align*}
  \vec{b}^{x} &=
  \Big[ \textstyle
    \sum_i c_i p^\star_{i10} \;\;\;
    \sum_i c_i p^\star_{i20} \;\;\;
    \cdots \;\;\;
    \sum_i c_i p^\star_{iN0}
  \Big]^T,
\\
  \vec{b}^{y} &=
  \Big[ \textstyle
    \sum_j c_j p^\star_{1j0} \;\;\;
    \sum_j c_j p^\star_{2j0} \;\;\;
    \cdots \;\;\;
    \sum_j c_j p^\star_{Nj0}
  \Big]^T,
\end{align*}
where $\mat{I}$ is the $N \times N$ identity matrix and $\mat{C}^{(1)}$ and $\mat{C}^{(2)}$ are a $N \times N$ sparse matrices that only depend on the parameters $c_i$'s.
The vectors of incoming characteristics, $\vec{r}^{-x}$ and $\vec{r}^{-y}$, contain the unknowns of system \eqref{eqn:systEdgeSymmetry}, while $\vec{r}^{+x}$, $\vec{r}^{+y}$, $\vec{b}^x$ and $\vec{b}^y$ are computed using the fields at $t=t_2$.
Because the matrix of the system (\ie the matrix in the left-hand side) is the same for all the edges, we precompute and store its inverse, after which we only need to do matrix-vector multiplications at each time step.
The matrix is composed of four $N \times N$ matrices, only two of which are independent.
Since its inverse has the same structure, only two $N \times N$ matrices must then be stored in memory.
Similarly, for the corner conditions, the inverse matrix of the $3N^2 \times 3N^2$ system \eqref{eqn:proc:cornerBC1}-\eqref{eqn:proc:cornerBC3} can be precomputed and stored.
It is composed of nine $N^2 \times N^2$ matrices, only three of which are independent and must be stored.

\end{enumerate}

This procedure is similar to the one described by Hagstrom and Warburton \cite{hagstrom2004new} for 2D cases, but with a different choice for the planar HABC.
Their formulation has been extended to a family of more general HABC, the complete radiation boundary conditions (CRBC) \cite{hagstrom2009complete,hagstrom2010radiation}, which can be accurate for both traveling and evanescent waves.
With a specific choice of parameters, corresponding to the Pad\'e case, these HABCs are equivalent to the one used here, but the formulations are written differently.

With the HABC of Hagstrom and Warburton in the Pad\'e case, fields from all the levels appear in the compatibility condition at corners.
This leads to an inconsistent formulation when using discontinuous Galerkin schemes based on an unstructured mesh.
Indeed, both 3D and 2D fields can have more than one value at the corners of the domain if several tetrahedral or triangular mesh cells touch this corner.
The same inconsistency appears when deriving the compatibility conditions with the HABC proposed by Collino \cite{collino1993conditions,collino1993high}.
By contrast, with our formulation, the 0D fields $p_{ijk}$ are defined only with 1D fields (equation \eqref{eqn:lev3:ijkCorner}), which have only one value at corners.
This observation motivated our choice for the HABC specifically based on the approximate square root represented by equation \eqref{eqn:approxRoot2}.

\subsection{Discontinuous Galerkin time domain scheme}
\label{sec:scheme:DG}

The HABC and compatibility conditions are discretized using a nodal discontinuous Galerkin finite element method with upwind fluxes in space and a low-storage fourth-order Runge-Kutta method in time \cite{hesthaven2007nodal}.
In the complete boundary procedure for cuboidal domains, the pressure-velocity system must be solved on the edges, on the faces and in the volume of the domain, which leads to a multidimensional solver.
We have used the 1D, 2D and 3D versions of the same spatial scheme.

The cuboidal domain is partitioned into a volume mesh of $K^\text{tet}$ non-overlapping tetrahedral cells, $\Omega = \bigcup_k \D_k^\text{tet}$, where $\D_k^\text{tet}$ is the $k^\text{th}$ cell.
Surface and line meshes are built on this volume mesh: the surface mesh is composed of the cell faces belonging to faces of the domain where the HABC is prescribed, while the line mesh is composed of the cell edges belonging to edges of the domain where two HABCs cross.
We denote by $K^\text{tri}$ the number of triangular cells $\D_k^\text{tri}$, by $K^\text{lin}$ the number of line cells $\D_k^\text{lin}$, and by $K^\text{pnt}$ the number of corners where three HABCs cross.
For instance, we have $K^\text{pnt}=8$ if the original problem is defined on the infinite space $\mathbb{R}^3$, and $K^\text{pnt}=4$ if a homogeneous Dirichlet boundary condition is prescribed on one face of the domain.
The volume mesh has $F_\bnd^\tet=K^\text{tri}$ boundary cell faces where a HABC are prescribed.
Since, an a cube, each edge has two neighboring faces and each corner has three neighboring edges, we assume that the surface and line meshes have respectively $F_\bnd^\tri=2K^\text{lin}$ and $F_\bnd^\lin=3K^\text{pnt}$ boundary cell faces where a HABC are prescribed.

For each of the 1D, 2D and 3D solvers, the pressure fields and the Cartesian components of the velocity fields are approximated by piecewise polynomial functions, which are discontinuous at the interface between two cells.
The discrete unknowns correspond to the values of fields at nodes distributed over the boundary and the interior of an element \cite{hesthaven2007nodal}.
In this work, the spatial distribution of nodes in the reference tetrahedron is defined using the \textit{Warp \& Blend} technique \cite{warburton2006explicit}.
The nodes in the reference triangle and on the reference line are chosen to match the face nodes and the edge nodes, respectively, of the tetrahedron.

The spatial scheme is built on a variational form of the equations.
Hereafter, the material properties $\rho$ and $c$ are assumed to be constant over each cell, but potentially discontinuous at the interfaces.
For each line, triangular and tetrahedral cell $\D_k$, we consider the variational form
\begin{align}
     \int_{\D_k} \dvp{p}{t}\:\psi \:d\vec{x}
   + \int_{\D_k} \rho c^2\:(\nabla\cdot\vec{u}) \: \psi \:d\vec{x}
   + \int_{\partial \D_k} (\rho c^2)^{\text{int}} \: \left((\vec{n}\cdot\vec{u})^\star-(\vec{n}\cdot\vec{u})^{\text{int}}\right) \: \psi \: d\vec{x}
  &= 0, \label{eqn:weakDG1}
  \\
     \int_{\D_k} \dvp{\vec{u}}{t}\cdot\boldsymbol{\psi}\:d\vec{x}
   + \int_{\D_k} \frac{1}{\rho} \: (\nabla p)\cdot\boldsymbol{\psi} \:d\vec{x}
   + \int_{\partial \D_k} \frac{1}{\rho^{\text{int}}} \: (p^\star-p^{\text{int}}) \: (\vec{n}\cdot\boldsymbol{\psi}) \:d\vec{x}
  &= 0, \label{eqn:weakDG2}
\end{align}
where $\psi(\vec{x})$ and $\boldsymbol{\psi}(\vec{x})$ are test functions, $\partial\D_k$ is the cell boundary and $\vec{n}$ is the outward unit normal to $\partial\D_k$.
The boundary conditions are prescribed and the solutions at the interface between two cells are coupled by selecting specific values for the numerical fluxes $p^\star$ and $(\vec{n}\cdot\vec{u})^\star$ in the boundary integrals of both equations.
At the interface between two elements, we consider the classical upwind fluxes provided by the exact Riemann solver \cite{wilcox2010high,hesthaven2007nodal,leveque2002finite},
\begin{align*}
  p^\star   &= \frac{\left\{p/(\rho c)\right\} - \vec{n}\cdot\jump{\vec{u}}}{\left\{1/(\rho c)\right\}}, \\
  \left(\vec{n}\cdot\vec{u}\right)^\star &= \frac{\vec{n}\cdot\left\{\rho c \vec{u}\right\} - \jump{p}}{\left\{\rho c\right\}},
\end{align*}
where $\{X\}=(X^{\text{ext}}+X^{\text{int}})/2$ and $\jump{X}=(X^{\text{ext}}-X^{\text{int}})/2$ are the average and the semi-jump, respectively, of any scalar or vector $X$.
The superscripts $\:^{\text{ext}}$ and $\:^{\text{int}}$ denote the exterior and interior values at the interface.
If the medium is homogeneous at the interface, the numerical fluxes can be conveniently rewritten as
\begin{align}
  p^\star &= (r^+)^{\text{int}} + (r^-)^{\text{ext}}, \label{eqn:fluxDG1} \\
  \left(\vec{n}\cdot\vec{u}\right)^\star &= \frac{(r^+)^{\text{int}} - (r^-)^{\text{ext}}}{\rho c}, \label{eqn:fluxDG2}
\end{align}
where $r^+$ and $r^-$ are the outgoing and incoming characteristics, respectively, defined as
\begin{align*}
  r^+ &= p + \rho c \: (\vec{n}\cdot\vec{u}), \\
  r^- &= p - \rho c \: (\vec{n}\cdot\vec{u}).
\end{align*}
At the domain boundary, the basic ABC is straightforwardly incorporated in the formulation by using the numerical fluxes \eqref{eqn:fluxDG1}-\eqref{eqn:fluxDG2} with the incoming characteristic equal to zero.
In the HABC procedure, boundary conditions for the 3D, 2D and 1D solvers are enforced by defining incoming characteristics using the 2D, 1D and 0D solvers, respectively.
Finally, the homogeneous boundary condition $p=0$ is enforced by taking
\begin{align*}
  p^\star &= 0, \\
  \left(\vec{n}\cdot\vec{u}\right)^\star &= \vec{n}\cdot\vec{u}^{\text{int}} + \frac{p^{\text{int}}}{(\rho c)^{\text{int}}}.
\end{align*}

For each element $\D_k$, the semi-discrete equations are obtained by substituting the semi-discrete fields into the variational form \eqref{eqn:weakDG1}-\eqref{eqn:weakDG2}, and using the Lagrange polynomials as test functions \cite{hesthaven2007nodal}.
For each field, this leads to a system that reads
\begin{align*}
  \frac{d\vec{q}_k}{dt} = \vec{r}_k,
\end{align*}
where the vectors $\vec{q}_k$ and $\vec{r}_k$ contain the discrete unknowns and the values of the right-hand side  terms for $\D_k$.
The right-hand side vector can be written as
\begin{equation}
  \vec{r}_{k}
    = \sum_{i=1}^{N_\text{dim}} \sum_{j=1}^{N_\text{dim}}
      g_{k,i,j}^{\text{vol}} \: \vec{D}_{j} \: \vec{f}_{k,j}
    + \sum_{f=1}^{N_\text{faces}}
      g_{k,f}^{\text{sur}} \: \vec{L}_{f} \: \vec{p}_{k,f},
  \label{eqn:dgLocalRhs}
\end{equation}
where $N_\text{dim}$ is the spatial dimension of the element, $N_\text{faces}$ is the number of faces, $\vec{f}_{k,j}$ corresponds to the physical flux in the $x_j$-direction for all the nodes of $\D_k$, and the vector $\vec{p}_{k,f}$ contains the boundary term for all the nodes belonging to the face $f$.
In the right-hand side vector \eqref{eqn:dgLocalRhs}, the first term (called the \textit{volume term}) corresponds to the integrals over the cell $\D_k$, and the second term (called the \textit{surface term}) corresponds to those over its boundary $\partial\D_k$.
The matrices $\mat{D}_{j}$ and $\mat{L}_{f}$ are respectively differentiation and lifting matrices for the reference element, while the geometric factors $g_{k,i,j}^{\text{vol}}$ and $g_{k,f}^{\text{sur}}$ depend on the shape of each element.
The matrices and factors are defined in \cite{hesthaven2007nodal}.
The semi-discrete equations are explicitly derived in \cite{modave2015nodal} for the three-dimensional case.

The low-storage fourth-order Runge-Kutta scheme is used for time discretization.
This scheme has five stages and require the storage of an auxiliary residual vector $\vec{s}_k$.
At each stage $n$ of each time iteration $m$, the residual vector $\vec{s}_k^{m+n/5}$ and the unknown vector $\vec{q}_k^{m+n/5}$ are updated according to
\begin{align}
  \vec{s}_k^{m+n/5} &= a_n \: \vec{s}_k^{m+(n-1)/5} + \vec{r}_k^{m+(n-1)/5}, \label{eqn:timeStepProc1} \\
  \vec{q}_k^{m+n/5} &= \vec{q}_k^{m+(n-1)/5} + b_n \: \vec{s}_k^{m+n/5},     \label{eqn:timeStepProc2}
\end{align}
where $\vec{r}_k^{m+n/5}$ and $\vec{s}_k^{m+n/5}$ correspond to the vector computed at time $t=(m+c_i n)\Delta t$.
The values of the coefficients $a_n$, $b_n$ and $c_n$ can be found in \citep{carpenter1994fourth}.

\subsection{GPU-accelerated computational implementation}
\label{sec:implementation}

We have implemented the boundary procedure in a discontinuous Galerkin code programmed using the \CC~language with the OCCA library \cite{medina2014occa} for GPU computing.
Discontinuous finite element schemes have attractive features for parallel computing on multi-threading devices such as GPU, but a careful implementation is required to optimize the efficiency of the solver (see \eg \cite{klockner2009nodal,modave2016gpu,fuhry2014discontinuous}).
In order to improve the computational efficiency, implementation strategies have been studied for advanced discontinuous Galerkin schemes with hybrid meshes \cite{chan2016gpu,chan2016reduced}, Bernstein-Bezier basis functions \cite{chan2015gpu}, multi-rate time-stepping schemes \cite{godel2010gpu,gandham2015gpu} and distributed parallel computing on GPU clusters \cite{modave2015nodal} in several application contexts.
In this work, we propose a single-GPU implementation based on the nodal discontinuous Galerkin method and the  time-stepping scheme presented in the previous section.
We highlight that the implementation strategies used here are compatible with those presented in the above references.

The 3D solver is implemented following strategies described in \cite{modave2015nodal,modave2016gpu}.
We have implemented the 1D and 2D solvers in a similar way, with the supplementary tasks required for the boundary procedure.
A specific implementation has been conceived for the 0D solver, which only solves the compatibility system at corners using the inverse matrix of this system.

\subsubsection*{Memory management}

All the data required for computation are stored in the global memory of the GPU.
For each of the 1D, 2D and 3D solvers, a floating-point array \texttt{q} stores all the discrete unknowns of the solver, while the array \texttt{qf} contains a copy of traces associated to face nodes (\ie $p$ and $\vec{n}\cdot\vec{u}$, where $\vec{n}$ is the outward unit normal to the face).
The arrays \texttt{rhs} and \texttt{res} store the right-hand side terms and the residual, respectively, used for the time-stepping procedure \eqref{eqn:timeStepProc1}-\eqref{eqn:timeStepProc2}.
The array \texttt{qb} contains the incoming characteristic variables used as boundary condition at boundary nodes of each mesh.
Arrays are used to store the elemental matrices (\texttt{Drst} and \texttt{Lift}) and the geometric and physical parameters (\texttt{volPar} and \texttt{surPar}) required to compute the right-hand side terms \eqref{eqn:dgLocalRhs}.
Additional arrays used by the 0D, 1D and 2D solvers for the boundary procedure are allocated: the HABC coefficients (\texttt{coefHabc}), the inverse matrices of the compatibility systems (\texttt{matHabcEdge} and \texttt{matHabcCorner}) and a temporary storage \texttt{q0} (its purpose is explained later).

An array of integers, the connectivity array \texttt{map}, is used when computing the numerical fluxes at face nodes in the surface kernel.
This array has one entry for each face node: a positive value gives the address of the corresponding face node on the neighboring cell  in \texttt{qf}, a negative value corresponds to a characteristic-based boundary condition and gives the address of the incoming characteristic in \texttt{qb}, and a zero value corresponds to an homogeneous Dirichlet boundary condition.
Connectivity arrays between the nodes and faces nodes of meshes with different spatial dimensions are also used.

The granularity of storage of all the arrays has been chosen in order to maximize coalescing transfers and data reuse (see \eg \cite{klockner2009nodal,modave2016gpu}).
The main parameters of the solvers are defined in table \ref{tab:impHabc:params}.
The sizes and granularity of the main arrays are given in table \ref{tab:impHabc:arrays}.

\begin{table}[!t]
\centering
\small
\begin{tabular}{l|l|c|c|c} \hline
  Definition & Symbol & 1D & 2D & 3D \\ \hline
  Number of elements in the mesh       & $K$              & $K^\text{lin}$ & $K^\text{tri}$ & $K^\text{tet}$ \\
  Number of HABC boundary faces        & $F_\text{bnd}$   & $3K^\text{pnt}$ & $2K^\text{lin}$ & $K^\text{tri}$ \\
  Spatial dimension           & $N_\text{dim}$   & 1 & 2 & 3 \\
  Number of faces per element & $N_\text{faces}$ & 2 & 3 & 4 \\
  Number of nodes per element & $N_p$    &\small $P+1$ &\small $(P+1)(P+2)/2$ &\small $(P+1)(P+2)(P+3)/6$ \\
  Number of nodes per face    & $N_{fp}$ &\small $1$   &\small $P+1$          &\small $(P+1)(P+2)/2$ \\
  Number of (scalar) fields   & $N_\text{fields}$ & 2 & 3 & 4 \\
  Number of traces            & $N_\text{traces}$ & 2 & 2 & 2 \\
  Number of set of fields     & $N_\text{sets}$ & $N^2$ & $N$ & $1$ \\ \hline
\end{tabular}
\caption{Definition and values of the differents parameters used in the multidimensional implementation.
$P$ is the polynomial degree of the basis functions.
$N$ is the number of auxiliary fields in the planar HABC.}
\label{tab:impHabc:params}
\end{table}

\begin{table}[!t]
\centering
\small
\begin{tabular}{l|l|l} \hline
  Definition & Symbol & Size \\ \hline
    Unknown fields at nodes & \texttt{q} & $K \cdot N_\text{sets} \cdot N_\text{fields} \cdot N_p$ \\
    Unknown traces at face nodes & \texttt{qf} & $K \cdot N_\text{sets} \cdot N_\text{traces} \cdot N_\text{faces} \cdot N_{fp} $ \\
    Incoming characteristics at boundary face nodes & \texttt{qb} & $F_\text{bnd} \cdot N_\text{sets} \cdot N_{fp} $ \\
    Right-hand side array & \texttt{rhs} & $K \cdot N_\text{sets} \cdot N_\text{fields} \cdot N_p$ \\
    Residual array & \texttt{res} & $K \cdot N_\text{sets} \cdot N_\text{fields} \cdot N_p$ \\
    Differentiation matrices ($\mat{D}_1$, $\mat{D}_2$, \dots) & \texttt{Drst} & $N_p^2 \cdot N_\text{dim}$ \\
    Lifting matrices ($\mat{L}_1$, $\mat{L}_2$, \dots) & \texttt{Lift} & $N_\text{faces} \cdot N_{fp} \cdot N_p$ \\
    HABC coefficients $c_i$'s &  \texttt{coefHabc} & $N$ \\
    Inverse matrix of the compatibility system at edges & \texttt{matHabcEdge}    & $2\cdot N^2$ \\
    Inverse matrix of the compatibility system at corners & \texttt{matHabcCorner} & $3\cdot N^4$ \\ \hline
\end{tabular}
\caption{Main arrays stored in the global memory of the GPU for the 1D, 2D and 3D solvers.
The last three arrays are used by the 0D, 1D and 2D solvers.
Sizes of arrays are written from the coarsest to the finest granularity of storage.
The symbols are defined in table \ref{tab:impHabc:params}.
}
\label{tab:impHabc:arrays}
\end{table}

\subsubsection*{Kernels}

The computational procedure is decomposed into several subtasks implemented in separate OCCA kernels.
This allows us to optimize each task considering the properties of both the task and the GPU.
Our implementation has three main kernels for each of the 1D, 2D, 3D solvers:
\begin{enumerate}
\setlength\itemsep{0em}
  \item the \textit{volume kernel} computes the first term of the right-hand side vector \eqref{eqn:dgLocalRhs};
  \item the \textit{surface kernel} computes the second term of the right-hand side vector \eqref{eqn:dgLocalRhs};
  \item the \textit{update kernel} performs the time stepping (equations \eqref{eqn:timeStepProc1}-\eqref{eqn:timeStepProc2}) and updates the incoming and outgoing characteristics by performing the operations listed in algorithm \ref{algo:multiSolver}.
\end{enumerate}
For the 0D solver, there is a single \textit{update kernel}, which updates the incoming characteristics used by 1D solver.
All the kernels are called at each stage of each time step in a specific order: first the three volume kernels (in any order), then the three surface kernels (in any order), and finally the four update kernels (starting with the 3D and ending with the 0D, following the procedure in algorithm \ref{algo:multiSolver}).

The volume and surface kernels consist of streaming operations and element-wise matrix-vector multiplications.
In a nutshell, the volume kernels load the values of fields from \texttt{q} for each element, compute the physical fluxes at each node, perform the matrix-vector products using \texttt{Drst}, and store the result in \texttt{rhs}.
The surface kernels load the values of traces and incoming characteristics from \texttt{qf} and \texttt{qb} for each element, compute the numerical fluxes at each face node, perform the matrix-vector products using \texttt{Lift}, and update \texttt{rhs} with the result.

All the volume and surface kernels are written and optimized in a similar way.
In the GPU programming model, a \textit{thread} is the smallest sequence of instructions that are managed independently with their own private memory.
Threads belonging to the same \textit{thread block} run concurrently and can collaborate using shared memory.
Following \cite{klockner2009nodal}, the tasks of the volume and surface kernels are parallelized by associating one thread to the computational work required for one node, and by associating one thread block to several elements.
In the volume and surface kernels, $N_p$ and $\max(N_p, N_{\text{faces}} N_{fp})$ threads are dedicated to one element, and one thread block is dedicated to $K_\text{blkV}$ and $K_\text{blkS}$ elements, respectively.
The parameters $K_\text{blkV}$ and $K_\text{blkS}$ provide a way to tune the occupation of the GPU for each kernel of each solver.
The 3D kernels and further details about the optimization strategies can be found in \cite{modave2015nodal}.
In the 1D and 2D kernels, several sets of fields ($N^2$ and $N$, respectively) are associated to each node, and the operations are performed several times with the different fields, still associating one node per thread.
Since the elemental matrices and parameters are identical for each set of fields, the kernels are written to enable reuse of these data.
Aside from this difference, the 1D and 2D kernels are similar to the 3D kernels.

\begin{algorithm}[!tb]
\caption{3D update kernel}
\label{algo:updateKernel3}
\vspace{0.1cm}
\textbf{input} \\
\hspace{4mm} \textit{pointers} $^*\texttt{q}^\tet$, $^*\texttt{qf}^\tet$, $^*\texttt{rhs}^\tet$ and $^*\texttt{res}^\tet$ \;

\ParFor{\normalfont each block $b$ of elements}{

\textit{shared array} \texttt{val} \ {\color{gray}(array $N_\fields^\tet \cdot K_\blkU^\tet \cdot N_{p}^\tet$)} \;

\ParFor{\normalfont each element $k$ of block $b$}{
\ParFor{\normalfont each node $n$ of element $k$}{
    compute updated 3D residuals using \texttt{rhs}$^\tet$ and \texttt{res}$^\tet$, and store in \texttt{val} \;
        \quad $\rightarrow$ store in \texttt{val}, then save in \texttt{res}$^\tet$ \;
    compute updated 3D fields $(p,\vec{u})$ using \texttt{q}$^\tet$ and \texttt{val}, and store in \texttt{val} \;
        \quad $\rightarrow$ store in \texttt{val}, then save in \texttt{q}$^\tet$ \;
}
}

\textbf{memory fence}

\ParFor{\normalfont each element $k$ of block $b$}{
\ParFor{\normalfont each face node $n_f$ of element $k$}{
    compute updated 3D traces using \texttt{val} \;
      \quad $\rightarrow$ save in \texttt{qf}$^\tet$ \;
}
}

}
\vspace{0.1cm}
\end{algorithm}

\begin{algorithm}[!tb]
\caption{2D update kernel}
\label{algo:updateKernel2}
\vspace{0.1cm}
\textbf{input} \\
\hspace{4mm} \textit{pointers} $^*\texttt{q}^\tri$, $^*\texttt{qf}^\tri$, $^*\texttt{rhs}^\tri$ and $^*\texttt{res}^\tri$\;
\hspace{4mm} \textit{pointers} $^*\texttt{qf}^\tet$, $^*\texttt{qb}^\tet$, $^*\texttt{q0}^\tet$, $^*\texttt{parHabc}$\;

\ParFor{\normalfont each block $b$ of elements}{

\textit{shared array} \texttt{val} \ {\color{gray}(array $N_\fields^\tri \cdot N \cdot K_\blkU^\tri \cdot N_{p}^\tri$)} \;

\ParFor{\normalfont each element $k$ of block $b$}{
\ParFor{\normalfont each node $n$ of element $k$}{
    \textit{private float} \texttt{charIn}, \texttt{charOut}, \texttt{charSum} \;
    \For{\normalfont each 2D set $s\in\{1,\dots,N\}$}{
      compute updated 2D residuals using \texttt{rhs}$^\tri$ and \texttt{res}$^\tri$ \;
        \quad $\rightarrow$ store in \texttt{val}, then save in \texttt{res}$^\tri$ \;
    }
    load the previous sum of 3D characteristics from \texttt{q0}$^\tet$ to \texttt{charSum} \;
    \For{\normalfont each 2D set $s\in\{1,\dots,N\}$}{
      compute previous 2D fields $p^\star$ using \texttt{q}$^\tri$ and \texttt{charSum} \;
      compute updated 2D fields $(p^\star,\vec{u})$ using \texttt{val} and \texttt{q}$^\tri$ \;
        \quad $\rightarrow$ store in \texttt{val} \;
    }
    compute updated 3D outgoing characteristics using \texttt{qf}$^\tet$ \;
      \quad $\rightarrow$ store in \texttt{charOut} \;
    compute updated 3D incoming characteristics using \texttt{val}, \texttt{charOut} and \texttt{parHabc} \;
      \quad $\rightarrow$ store in \texttt{charIn}, then save in \texttt{qb}$^\tet$ \;
    compute updated sum of 3D characteristics using \texttt{charOut} and \texttt{charIn} \;
      \quad $\rightarrow$ store in \texttt{charSum}, then save in \texttt{q0}$^\tet$ \;
    \For{\normalfont each 2D set $s\in\{1,\dots,N\}$}{
      compute updated 2D fields $p$ using \texttt{val} and $\texttt{charSum}$ \;
      \quad $\rightarrow$ store in \texttt{val}, then save the updated 2D fields $(p,\vec{u})$ in \texttt{q}$^\tri$ \;
    }
}
}

\textbf{memory fence}

\ParFor{\normalfont each element $k$ of block $b$}{
\ParFor{\normalfont each face node $n_f$ of element $k$}{
    \For{\normalfont each 2D set $s\in\{1,\dots,N\}$}{
      compute updated 2D traces $(p,\vec{n}\cdot\vec{u})$ using \texttt{val} \;
      \quad $\rightarrow$ save in \texttt{qf}$^\tri$ \;
    }
}
}

}
\vspace{0.1cm}
\end{algorithm}

\begin{algorithm}[!tbp]
\caption{1D update kernel}
\label{algo:updateKernel1}
\vspace{0.1cm}
\textbf{input} \\
\hspace{4mm} \textit{pointers} $^*\texttt{q}^\lin$, $^*\texttt{qf}^\lin$, $^*\texttt{rhs}^\lin$ and $^*\texttt{res}^\lin$\;
\hspace{4mm} \textit{pointers} $^*\texttt{qf}^\tri$, $^*\texttt{qb}^\tri$, $^*\texttt{q0}^\tri$, $^*\texttt{parHabc}$, $^*\texttt{matHabcEdge}$\;

\ParFor{\normalfont each block $b$ of elements}{

\textit{shared array} \texttt{charSum}, \texttt{charOut}, \texttt{vecX} \ {\color{gray}(arrays $2 \cdot K_\blkU^\lin \cdot N \cdot N_{p}^\lin$)} \;
\textit{shared array} \texttt{val} \ {\color{gray}(array $N_\fields^\lin \cdot N \cdot K_\blkU^\lin \cdot N \cdot N_{p}^\lin$)} \;

\ParFor{\normalfont each element $k$ of block $b$}{
\ParFor{\normalfont each 2D set $s\in\{1,\dots,N\}$}{
\ParFor{\normalfont each node $n$ of element $k$}{
    \For{\normalfont each neighboring face $side\in\{1,2\}$}{
        load previous sum of 2D characteristics from \texttt{q0}$^\tri$ to \texttt{charSum} \;
        compute updated 2D outgoing characteristics using \texttt{qf}$^\tri$ \;
          \quad $\rightarrow$ store in \texttt{charOut} \;
    }
}
}
}

\textbf{memory fence}

\ParFor{\normalfont each element $k$ of block $b$}{
\ParFor{\normalfont each 1D set group $s'\in\{1,\dots,N\}$}{
\ParFor{\normalfont each node $n$ of element $k$}{
    \For{\normalfont each 1D set $s\in\{s',s'+N,\dots,s'+N\cdot(N-1)\}$}{
      compute updated 1D residuals using \texttt{rhs}$^\lin$ and \texttt{res}$^\lin$ \;
        \quad $\rightarrow$ store in \texttt{val}, then save in \texttt{res}$^\lin$ \;
    }
    \For{\normalfont each 1D set $s\in\{s',s'+N,\dots,s'+N\cdot(N-1)\}$}{
      compute previous 1D fields $p^\star$ using \texttt{q}$^\lin$, \texttt{charSum} and \texttt{parHabc} \;
      compute updated 1D fields $(p^\star,u)$ using \texttt{val} and \texttt{q}$^\lin$ \;
        \quad $\rightarrow$ store in \texttt{val} \;
    }
}
}
}

\textbf{memory fence}

\ParFor{\normalfont each element $k$ of block $b$}{
\ParFor{\normalfont each 2D set $s\in\{1,\dots,N\}$}{
\ParFor{\normalfont each node $n$ of element $k$}{
    \For{\normalfont each neighboring face $side\in\{1,2\}$}{
      compute right-hand side for the edge HABC system using \texttt{val}, \texttt{charOut}  and \texttt{parHabc} \!\!\!\!\!\!\!\!\! \;
        \quad $\rightarrow$ store in \texttt{vecX} \;
    }
}
}
}

\textbf{memory fence}

\ParFor{\normalfont each element $k$ of block $b$}{
\ParFor{\normalfont each 2D set $s\in\{1,\dots,N\}$}{
\ParFor{\normalfont each node $n$ of element $k$}{

    \For{\normalfont each neighboring face $side\in\{1,2\}$}{
      compute updated 2D incoming characteristics by using \texttt{matHabcEdge} and \texttt{vecX} \;
        \quad $\rightarrow$ save in \texttt{charIn}, then store in \texttt{qb}$^\tri$ \;
      compute sum of updated 2D characteristics using \texttt{charOut} and \texttt{charIn} \;
        \quad $\rightarrow$ store in \texttt{charSum}, then save in \texttt{q0}$^\tri$ \;
    }
}
}
}

\textbf{memory fence}

\ParFor{\normalfont each element $k$ of block $b$}{
\ParFor{\normalfont each 1D set group $s'\in\{1,\dots,N\}$}{
\ParFor{\normalfont each node $n$ of element $k$}{
    \For{\normalfont each 1D set $s\in\{s',s'+N,\dots,s'+N\cdot(N-1)\}$}{
      compute updated 1D fields $p$ using \texttt{val}, \texttt{charSum} and \texttt{parHabc} \;
      \quad $\rightarrow$ store in \texttt{val}, then save updated 1D fields $(p,u)$ in \texttt{q}$^\lin$ \;
    }
}
}
}

\textbf{memory fence}

\ParFor{\normalfont each element $k$ of block $b$}{
\ParFor{\normalfont each 1D set group $s'\in\{1,\dots,N\}$}{
\ParFor{\normalfont each face node $n_f$ of element $k$}{
    \For{\normalfont each 1D set $s\in\{s',s'+N,\dots,s'+N\cdot(N-1)\}$}{
      compute updated 1D traces using \texttt{val} \;
      \quad $\rightarrow$ save in \texttt{qf}$^\lin$ \;
    }
}
}
}

}
\vspace{0.1cm}
\end{algorithm}

\begin{algorithm}[!tb]
\caption{0D update kernel}
\label{algo:updateKernel0}
\vspace{0.1cm}
\textbf{input} \\
\hspace{4mm} \textit{pointers} $^*\texttt{qf}^\lin$, $^*\texttt{qb}^\lin$, $^*\texttt{parHabc}$, $^*\texttt{matHabcCorner}$ \;

\ParFor{\normalfont each element $k$}{

\textit{shared array} \texttt{charOut}, \texttt{vecX} \ {\color{gray}(arrays $3 \cdot N^2$)} \;

\ParFor{\normalfont each 1D set $s\in\{1,\dots,N^2\}$}{
    \For{\normalfont each neighboring edge $side\in\{1,2,3\}$}{
        compute updated 1D outgoing characteristics using \texttt{qf}$^\lin$ \;
          \quad $\rightarrow$ store in \texttt{charOut} \;
    }
}

\textbf{memory fence}

\ParFor{\normalfont each 1D set $s\in\{1,\dots,N^2\}$}{
    \For{\normalfont each neighboring edge $side\in\{1,2,3\}$}{
        compute right-hand side for the corner HABC system using \texttt{charOut} and \texttt{parHabc} \;
          \quad $\rightarrow$ store in \texttt{vecX} \;
    }
}

\textbf{memory fence}

\ParFor{\normalfont each 1D set $s\in\{1,\dots,N^2\}$}{
    \For{\normalfont each neighboring edge $side\in\{1,2,3\}$}{
        compute updated 1D incoming characteristics by using \texttt{matHabcCorner} and \texttt{vecX} \;
          \quad $\rightarrow$ save in \texttt{qb}$^\lin$ \;
    }
}

}
\vspace{0.1cm}
\end{algorithm}

The update kernels perform the time-stepping for the fields, and computes the incoming and outgoing characteristics following the boundary procedure in algorithm \ref{algo:multiSolver}.
The operations performed by the update kernels are rather different: the 3D kernel only updates fields (algorithm \ref{algo:updateKernel3}), the 2D kernel updates fields and performs streaming operations (algorithm \ref{algo:updateKernel2}), the 1D kernel updates fields and solves $2N\times2N$ systems (algorithm \ref{algo:updateKernel1}), the 0D kernel only solves $3N^2\times3N^2$ systems (algorithm \ref{algo:updateKernel0}).
The different kernels are written in algorithms \ref{algo:updateKernel3}-\ref{algo:updateKernel0} using pseudo-code to give an overview of the implementation.
In these algorithms, \textbf{parfor} denotes a parallel loop, while \textbf{for} denotes a sequential loop.
The most external parallel loop iterates over thread blocks, while the others iterate over threads.
Since the physical/geometrical factor arrays and the three connectivity arrays are used in a straightforward way, they are not mentioned in the algorithms for the sake of clarity.
We describe hereafter the key aspects of the kernels.

\begin{itemize}

\item
The 3D update kernel (algorithm \ref{algo:updateKernel3}) performs the time-stepping for the 3D fields in three steps: first the residual is updated at nodes with equation \eqref{eqn:timeStepProc1}, then the fields are updated at nodes with equation \eqref{eqn:timeStepProc2}, and finally the traces are computed at face nodes.
As for the volume and surface kernels, each thread deals with the tasks associated to a given node, and each thread block deals with $K_\text{blkU}^\tet$ elements.
There are therefore $K_\text{blkU}^\tet\cdot\max(N_p^\tet,N_\text{faces}^\tet\cdot N_{fp}^\tet)$ threads per thread block.

\item
The time-stepping of the 2D fields is performed by the 2D update kernel using the same parallelization strategy, with one node per thread and $K_\text{blkU}^\tri$ elements per thread blocks.
Since there are $N$ sets of 2D pressure and velocity fields per node, a sequential loop is used to iterate over the sets for each operation over these fields (algorithm \ref{algo:updateKernel2}).
In addition, the 3D incoming characteristic is computed sequentially by performing the operations described in algorithm \ref{algo:multiSolver}.

In this procedure, both the previous value and the updated value of the 3D outgoing characteristic are needed.
The first is used to compute the temporary 2D fields $p^\star$ at the beginning of the time step, and the second is used when updating the 3D incoming characteristic.
While the updated 3D outgoing characteristic is computed using the updated traces from array $\texttt{qf}^\tet$, we have introduced an additional array $\texttt{q0}^\tet$ that stores the sum of the 3D characteristics to compute $p^\star$ at the beginning of the time step.
This array is updated when the updated 3D incoming characteristic is available.

\item
The role of the 1D update kernel is similar to that of the 2D kernel: updating the 1D fields on the edges of the domain and computing the 2D incoming characteristics for the neighbor faces.
However, the size of data and the type of operations are different: there are $N^2$ 1D fields to update and $2N$ 2D incoming characteristics ($N$ for each neighbor face) to compute by performing matrix-vector products.

For this update kernel, we have modified the parallelization strategy in order to reduce the use of private memory storage and to perform matrix-vector products in parallel.
In the 1D kernel, $N$ threads deals with the tasks associated to one node.
Each of these threads is dedicated to $N$ sets of 1D fields and two 2D incoming characteristics.
In algorithm \ref{algo:updateKernel1}, there are therefore three inner parallel loops (over the elements, the sets and the nodes).
To preserve coalescing memory transfers, the inner most loop processes the nodes since they correspond to the finest granularity of storage for all the arrays.

As explained at the end of section \ref{sec:form:final}, the $2N \times 2N$ system to compute the 2D incoming characteristics can simply be done by performing four matrix-vector products with $N \times N$ matrices, where only two vectors and two matrices are different.
The 1D kernel computes the entries of both vectors (stored in a shared array \texttt{vecX}) and performs the matrix-vector products in a parallel way.
Each thread computes one entry of each vector, and computes one output value of each matrix-vector product.

\item
The only task of the 0D update kernel is computing the 1D outgoing characteristics by solving a $3N^2 \times 3N^2$ linear system for each corner, which the inverse matrix is available.
Using symmetry in the inverse matrix, the task consists in performing nine matrix-vector products with $N^2 \times N^2$ matrices, where only three vectors and three matrices are different.
The 0D kernel computes the entries of the three vectors (stored in a shared array \texttt{vecX}) and performs the matrix-vector products in a parallel fashion.
In our implementation, each thread block deals with one corner, and each thread computes one entry of each vector and one output value of each matrix-vector products.

\end{itemize}


\section{Numerical results}
\label{sec:results}

In this section, we present numerical results obtained with an academic benchmark (section \ref{sec:resu:valid}) and a realistic benchmark used in exploration geophysics (section \ref{sec:resu:seam}).

\subsection{Validation benchmark}
\label{sec:resu:valid}

In this section, the accuracy of the HABC formulation is studied with an academic benchmark which the solution is known.
We consider the propagation of a spherical wave in the infinite space $\mathbb{R}^3$.
The wave is generated with a source point at position $\vec{x}_\text{s}\in\Omega$ using the Ricker wavelet $s(t)$ defined as
\begin{align}
  s(t) \equiv \big(1-2\pi^2 f_\text{peak}^2(t-t_\text{s})^2\big) e^{-\pi^2 f_\text{peak}^2(t-t_\text{s})^2},
  \label{eqn:validBench:ricker}
\end{align}
where $f_\text{peak}$ is the peak frequency and $t_\text{s}$ is a time offset.
The source point is incorporated in the pressure-velocity system using a Dirac delta in the pressure equation:
\begin{align*}
  \dvp{p}{t} + \rho c^2 \nabla\cdot\vec{u} &= \delta(\vec{x}-\vec{x}_\text{s}) \:S(t),
\end{align*}
where $S(t)$ is the integral of the wavelet,
\begin{align*}
  S(t) = \int_{-\infty}^t s(t')\:dt' = (t-t_\text{s}) e^{-\pi^2 f_\text{peak}^2(t-t_\text{s})^2}.
\end{align*}
The analytic solution is obtained by taking the convolution of the Green function of the wave equation with the Ricker wavelet.
This solution then reads
\begin{subequations}
\begin{align}
  p^\text{ref}(t,\vec{x}) &= \frac{1}{4\pi r} s(t-r/c),
  \label{eqn:validBench:refSol1} \\
  \vec{u}^\text{ref}(t,\vec{x}) &= \frac{\vec{x}-\vec{x}_\text{s}}{4\pi \rho r^2}
    \left(\frac{1}{r} S(t-r/c) + \frac{1}{c} s(t-r/c) \right),
  \label{eqn:validBench:refSol2}
\end{align}
\end{subequations}
with $r=\|\vec{x}-\vec{x}_\text{s}\|^{1/2}$.


\subsubsection*{Setting}

The numerical simulation is performed on the cuboidal domain $\Omega = [-0.5,0.5]^3$ with a mesh composed of 70895 tetrahedra.
Third-degree polynomial basis functions are used (\ie $P=3$).
The time step $\Delta t$ is chosen according to
\begin{align}
  \Delta t = \mathrm{max}_k \frac{1}{c_k(P+1)^2 F_{\text{scale},k}},
  \label{eqn:validBench:timeStep}
\end{align}
where $c_k$ is the wave velocity and $F_{\text{scale},k}$ is the maximum ratio of surface to volume Jacobian of the $k^\text{th}$ mesh cell (see \eg \cite{warburton2003constants,chan2016gpu,hesthaven2007nodal}).
For this benchmark, we use dimensionless physical parameters $\rho_k$ and $c_k$ set to $1$.
The duration of the simulation is $t_\text{final}=5$ and the peak of the Ricker wavelet is generated at $t_\text{s}=0.5$ with the peak frequency $f_\text{peak}=2.5$.

We compare the numerical solution obtained with approximate boundary treatments (basic ABC and HABC) to the infinite-space reference solution \eqref{eqn:validBench:refSol1}-\eqref{eqn:validBench:refSol2}.
The difference is quantified with the relative $L^2$-error on the domain $\Omega$ defined as
\begin{align}
  \mathrm{Error}(t)
  = \sqrt{\frac{ \displaystyle
             \int_{\Omega} \left(
                \frac{1}{2\rho c^2} \left(p^{\text{ref}}(t,\vec{x})
                                         -p^{\text{num}}(t,\vec{x})\right)^2
              + \frac{\rho}{2} \left\|\vec{u}^{\text{ref}}(t,\vec{x})
                                     -\vec{u}^{\text{num}}(t,\vec{x})\right\|^2\right) d\vec{x}}{
           \text{Total energy generated by the source}}}.
  \label{eqn:validBench:error}
\end{align}
The total energy generated by the source (used in the denominator) is computed by performing the simulation with the homogeneous Dirichlet condition $p=0$ on $\partial\Omega$, which does not allow outgoing energy flux.
The energy then is evaluated in $\Omega$ when it reaches a constant value, that is when the wavelet is totally generated.
Both this energy and the error are evaluated using a numerical integration with a quadrature rule that is exact for seventh-degree polynomials.
Note that error \eqref{eqn:validBench:error} measures both modeling errors due to the approximate boundary treatment and numerical errors due to the discretization of the problem.


\subsubsection*{Results}

\begin{figure}[!t]
\centering
\begin{tabular}{cc}
\begin{subfigure}[b]{70mm}
  \centering
  \caption{Source centered $\vec{x}_\text{s}=(0,0,0)$}
  \includegraphics[width=68mm]{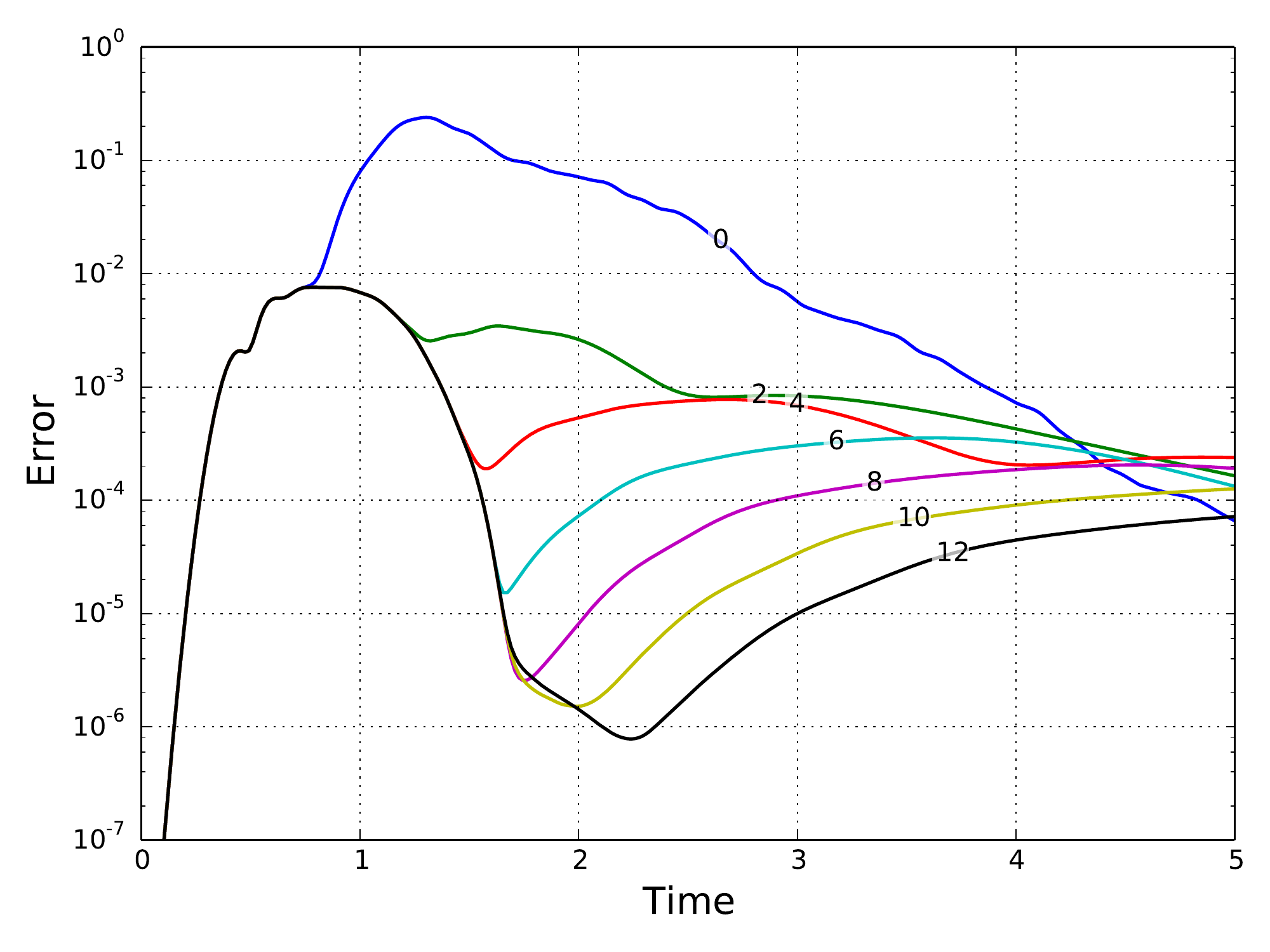}
  \label{fig:convBench:fullHabc:1}
\end{subfigure}
&
\begin{subfigure}[b]{70mm}
  \centering
  \caption{Source with offset $\vec{x}_\text{s}=(0.2,0.1,0)$}
  \includegraphics[width=68mm]{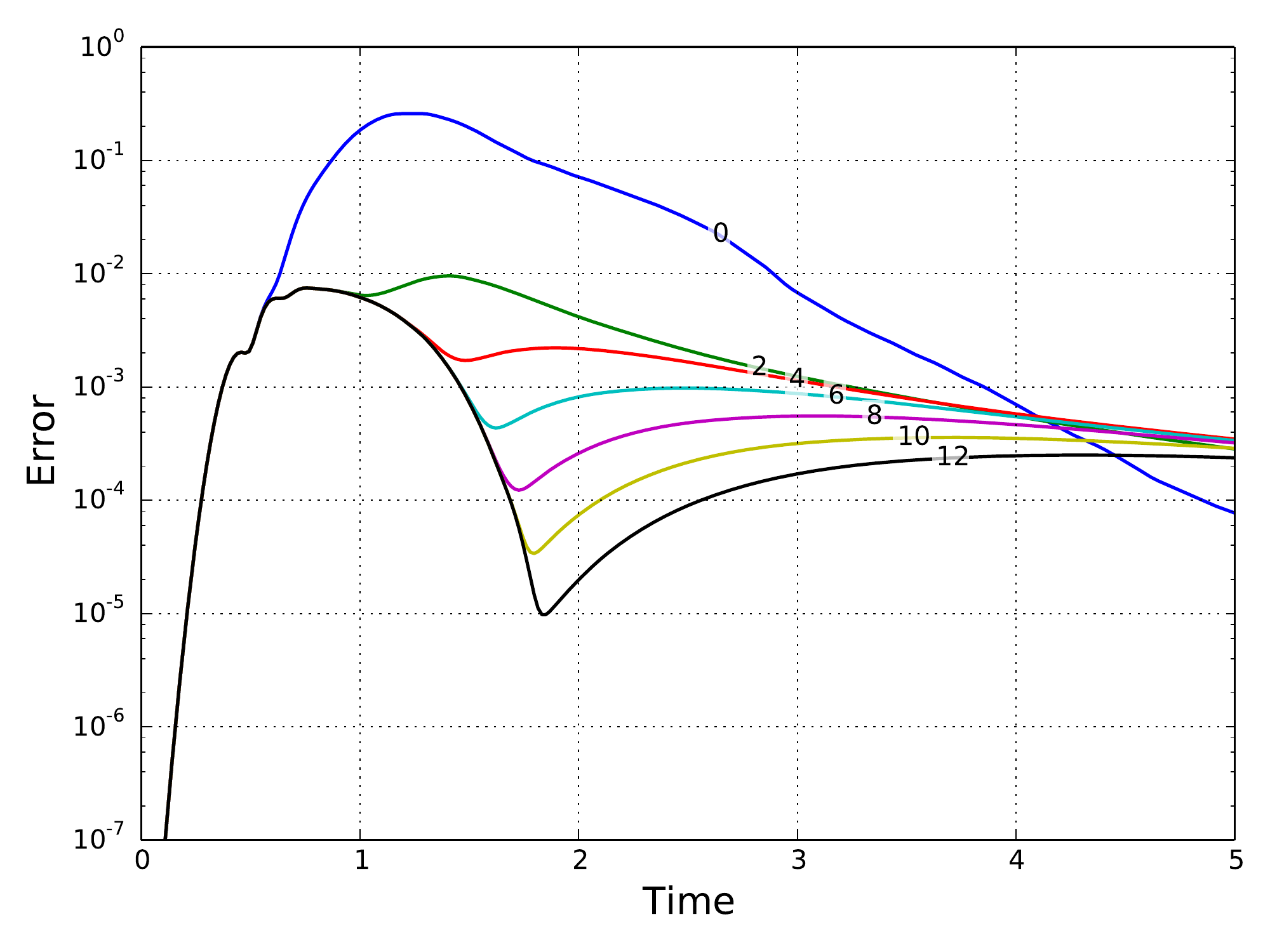}
  \label{fig:convBench:fullHabc:2}
\end{subfigure}
\end{tabular}
\vspace{-0.3cm}
\caption{Time-evolution of the error for the validation benchmark with approximate boundary treatments on all the faces of the domain.
The source is placed at the center of the domain (a) or slightly shifted away from the center (b).
The numbers on the curves indicate the number of additional fields in the boundary treatment ($N=0$ for the basic ABC and $N=2,4,\dots,12$ for HABC).}
\label{fig:convBench:fullHabc}
\end{figure}

Figure \ref{fig:convBench:fullHabc} shows the time-evolution of the error when the source is placed at the center of the domain ($\vec{x}_\text{s}=(0,0,0)$) and when it is slightly shifted ($\vec{x}_\text{s}=(0.2,0.1,0)$).
The generated spherical wavefront propagates in the domain and the peak reaches the boundary at $t=1$ and $t=0.8$, respectively.
At this instant, the error reaches $\sim 10^{-2}$ with all the boundary treatments and both sources.
In this first period, the total error is dominated by the numerical error.
After, the behavior depends on the boundary treatment.

With the basic ABC, the error increases until it reaches  $\sim 0.17$ for both sources, which means that approximately $17\%$ of the total energy generated in the domain has been reflected.
The error is clearly dominated by the modeling error: the reflected wavefront propagated in the domain is partially reflected at the boundary.
The error is continuously decreasing as the multiple reflections are absorbed.
With the basic ABC, it can be proved that, for both the continuous model and the numerical scheme, the energy cannot increases once the source is totally generated.
The observed error decay is therefore the expected result.

With the HABC, the error decreases until a minimum is reached between $t=1$ and $t=2.3$.
During this decrease, the error is the same with all the HABCs and is dominated by the numerical error.
The attained minimum depends on the order of the HABC: a larger order provides a smaller error.
We note that this minimum occurs earlier and is larger when the source is shifted (figure \ref{fig:convBench:fullHabc:2}).
Indeed, because the source is closer to a boundary, the reflection occurs earlier, and the amplitudes of both incident and reflected wavefronts are larger.
After the minimum, the error oscillates (for small $N$) or increases (for large $N$) to converge towards the same value for all orders of HABC.
Such phenomenon is well known with Pad\'e-like and Higdon-like boundary conditions (see the numerical results in \eg \cite{hagstrom2004new,hagstrom2007local,hagstrom2010radiation}).
It is due to the poor long time error behavior of these conditions, which can be overcome, for instance, with the CRBC \cite{hagstrom2009complete,hagstrom2010radiation}.

In order to validate HABC coupled with a homogeneous boundary condition, we consider a variant of the benchmark where $p=0$ is prescribed on the upper face (\ie $z=0.5$), while HABC are used on the other faces.
In this benchmark, a primary wavefront is generated by the source, and a secondary wavefront appears after the reflection of the primary front on the upper boundary.
The reference solution is straightforwardly obtained by using the infinite-space solution \eqref{eqn:validBench:refSol1}-\eqref{eqn:validBench:refSol2} with the method of images.

\begin{figure}[!t]
\centering
\begin{tabular}{cc}
\begin{subfigure}[b]{70mm}
  \centering
  \caption{Source centered $\vec{x}_\text{s}=(0,0,0)$}
  \includegraphics[width=68mm]{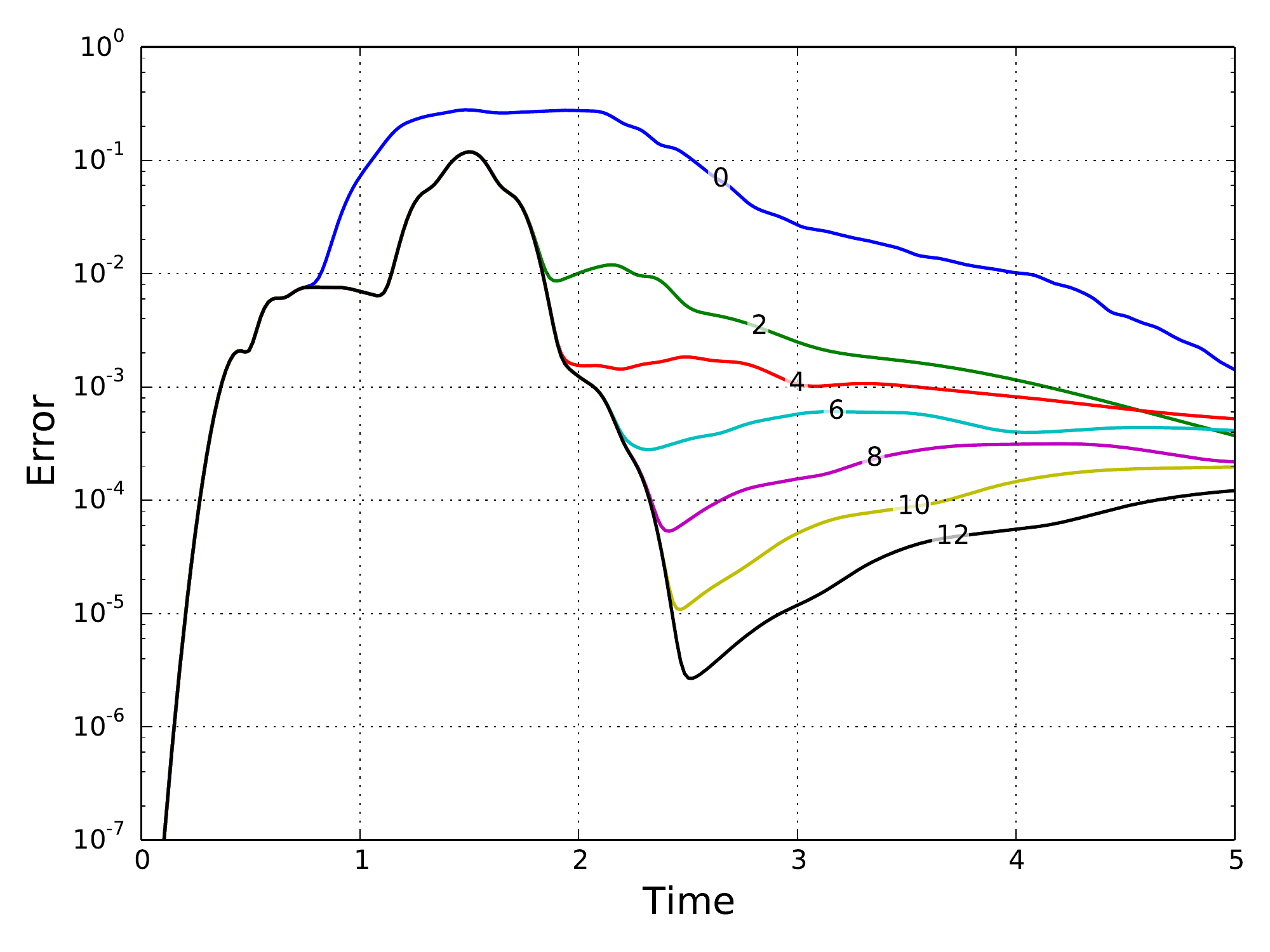}
  \label{fig:convBench:habcWithDirBC:1}
\end{subfigure}
&
\begin{subfigure}[b]{70mm}
  \centering
  \caption{Source with offset $\vec{x}_\text{s}=(0.2,0.1,0)$}
  \includegraphics[width=68mm]{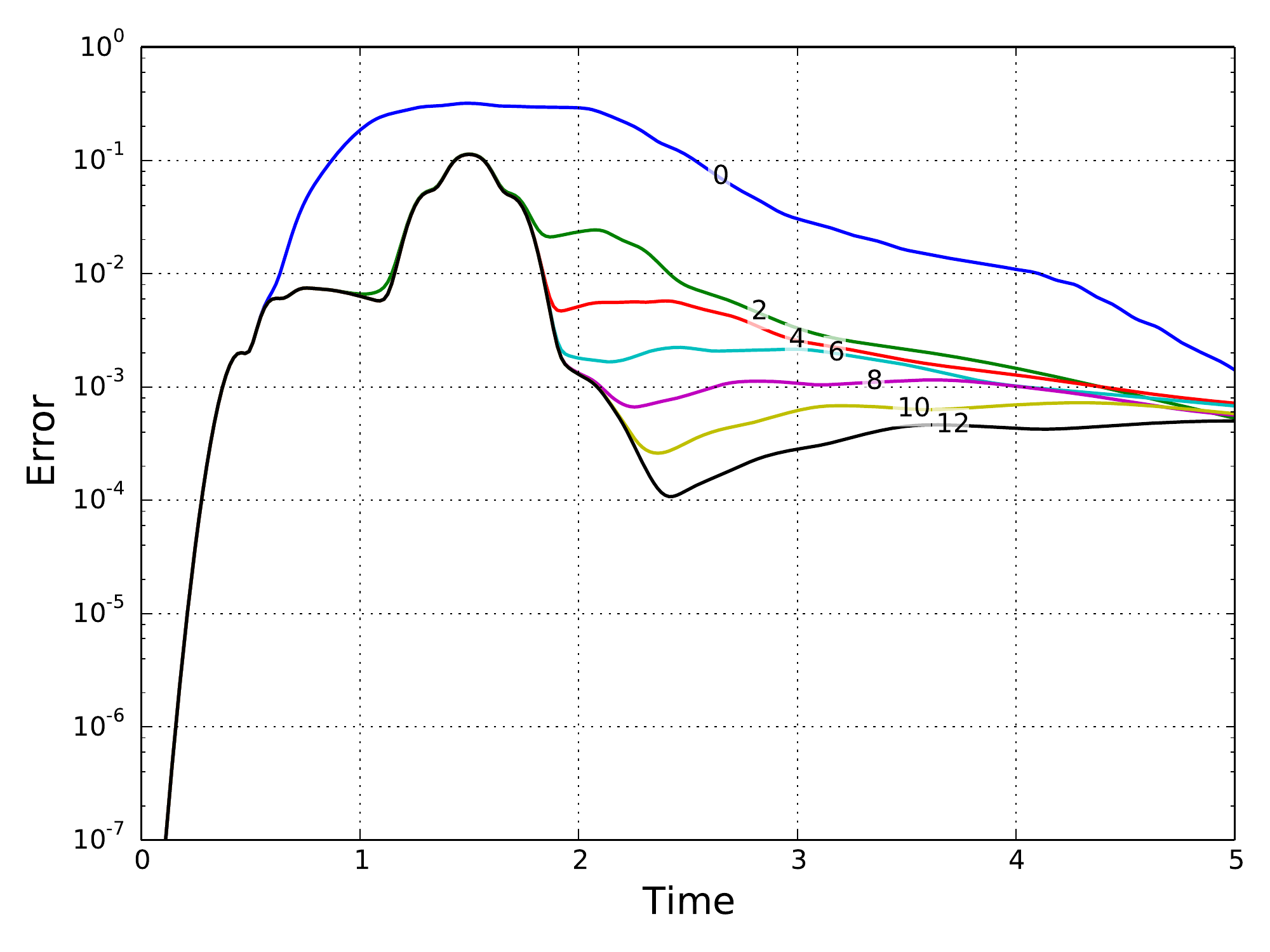}
  \label{fig:convBench:habcWithDirBC:2}
\end{subfigure}
\end{tabular}
\vspace{-0.3cm}
\caption{Time-evolution of the error for the validation benchmark with a free-surface boundary on the upper face of the domain (\ie $z=0.5$) and approximate boundary treatments on the other faces.
The source is placed at the center of the domain (a) or slightly shifted away from the center (b).
The numbers on the curves indicate the number of additional fields in the boundary treatment ($N=0$ for the basic ABC and $N=2,4,\dots,12$ for HABC).}
\label{fig:convBench:habcWithDirBC}
\end{figure}

Figure \ref{fig:convBench:habcWithDirBC} shows the time-evolution of the error for the modified benchmark.
The behavior of the error is the same as with the previous benchmark for $t<1$ because the second wavefront has not yet appeared.
In the range $t\in[1,2]$, all the HABC give the same error, which corresponds to the numerical error when both primary and secondary wavefronts are traveling in the domain.
With the basic ABC, the error is clearly dominated by modeling error due the spurious reflection of waves.
The long time error behavior is similar to the previous benchmark.



\subsection{Realistic benchmark}
\label{sec:resu:seam}

In order to test our approach with a more realistic situation, we have built a benchmark based on the \textit{SEAM Phase I} model produced by the \textit{SEG Advanced Modeling Program} \cite{fehler2011seam}.


\subsubsection*{Setting}

\begin{figure}[!t]
\centering
\begin{tabular}{cc}
\begin{subfigure}[b]{8cm}
  \centering
  \caption{Velocity model}
  \includegraphics[width=78mm]{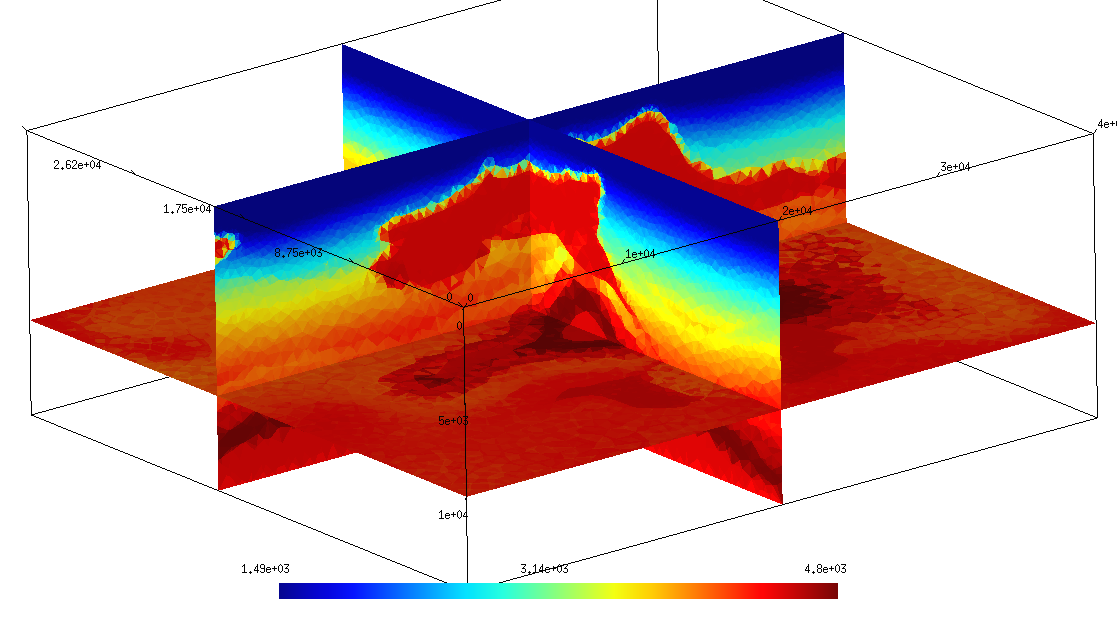}
  \label{fig:SEAM:model:V}
\end{subfigure}
&
\begin{subfigure}[b]{8cm}
  \centering
  \caption{Density model}
  \includegraphics[width=78mm]{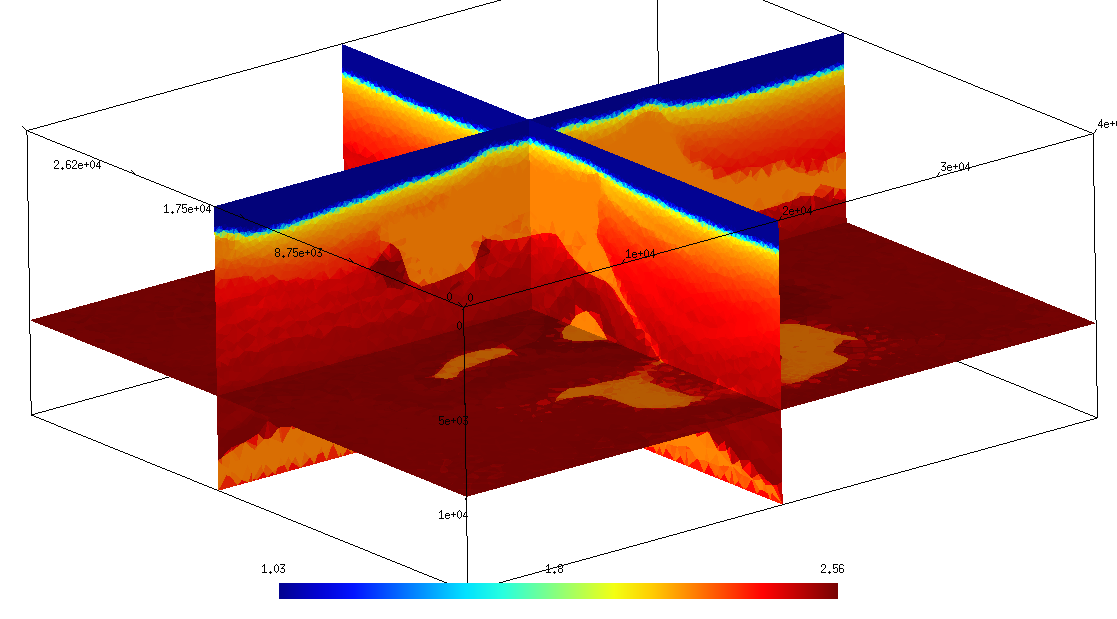}
  \label{fig:SEAM:model:D}
\end{subfigure}
\end{tabular}
\caption{Density $(a)$ and velocity $(b)$ based on the \textit{SEAM Phase I} model for the realistic benchmark.}
\label{fig:SEAM:model}
\end{figure}

\begin{figure}[!t]
  \centering
  \medskip
  \includegraphics[width=78mm]{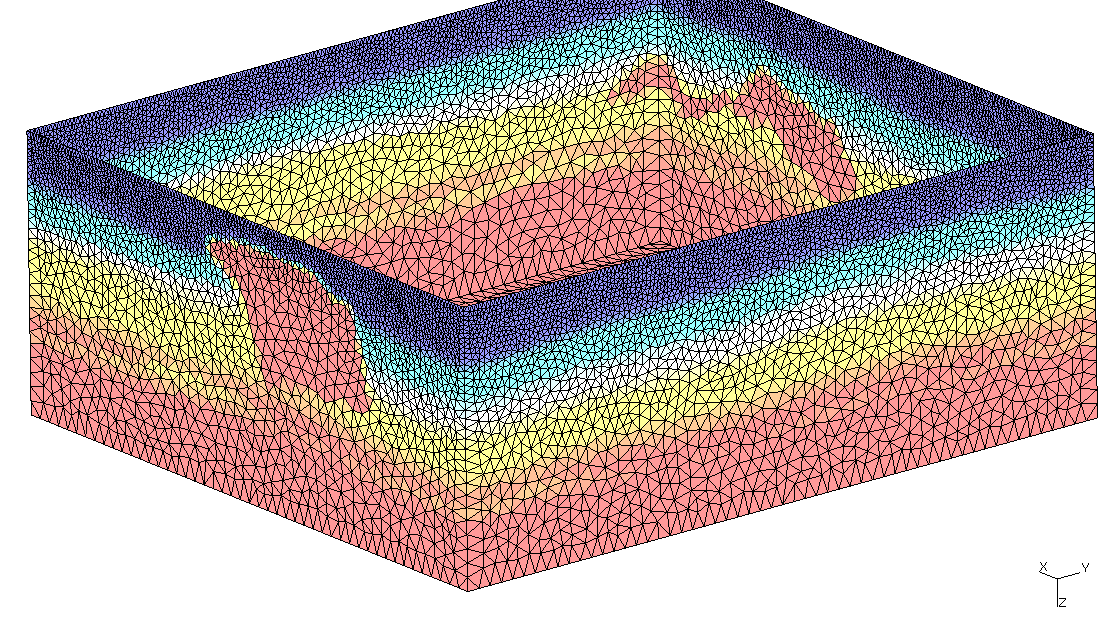}
\caption{Mesh based on the \textit{SEAM Phase I} model for the realistic benchmark.}
\label{fig:SEAM:mesh}
\end{figure}

The computational domain $\Omega = [0,35\km]\times[0,40\km]\times[0,15\km]$ of the \textit{SEAM Phase I} model is a 3D representation of a deepwater Gulf of Mexico salt domain with a stratigraphy.
The last dimension of the domain corresponds to the vertical direction.
The coordinate $z$ is the depth from the sea level.
The domain is partitioned into an unstructured mesh made of $1,179,989$ tetrahedra.
The mesh has been generated with smaller cells in regions with smaller P-wave velocity in order to accurately represent the spatial oscillations.
Since the physical parameters must be constant over each mesh cell, the mean density $\rho_k$ and the mean P-wave velocity $c_k$ of the SEAM model are taken.
The physical parameters and the mesh are shown on figures \ref{fig:SEAM:model} and \ref{fig:SEAM:mesh}.
A salt body is visible in the middle of the domain.
The upper part of the domain represents the ocean.

A wavefront is generated in the upper part of the domain by using a point source with the Ricker wavelet.
The position of the source is $\vec{x}_\text{s} = (10\km,10\km,1.5\km)$, the time offset is $t_\text{s}=2\s$ and the peak frequency is $f_\text{peak} = 2\Hz$.
The free-surface boundary condition is prescribed on the upper border of the domain (at $z=0$), while an HABC is used on the lateral and bottom borders.
For the boundary procedure, a surface mesh with $23,892$ triangles covers the lateral and bottom faces.
The line mesh composed of $332$ lines is used for the lateral and bottom edges.
Only the $4$ corners at the bottom are considered in the procedure.
The simulations have been performed with third-degree polynomial basis functions, for the duration $15\s$ with the global time-step $\Delta t = 0.124723\ms$, which has been computed using equation \eqref{eqn:validBench:timeStep}.

We solver this benchmark with one single Nvidia K40 GPU, which constrains the size of problems that can be solved.
For realistic applications, the solver can be improved by using strategies for parallel computing on GPU clusters and multi-rate time stepping \cite{modave2015nodal}.
Using these strategies dramatically accelerates the computation, and allows for larger and more refined meshes which support higher frequencies.



\subsubsection*{Qualitative comparison}

\begin{figure}[!p]
\begin{center}
  \includegraphics[width=160mm]{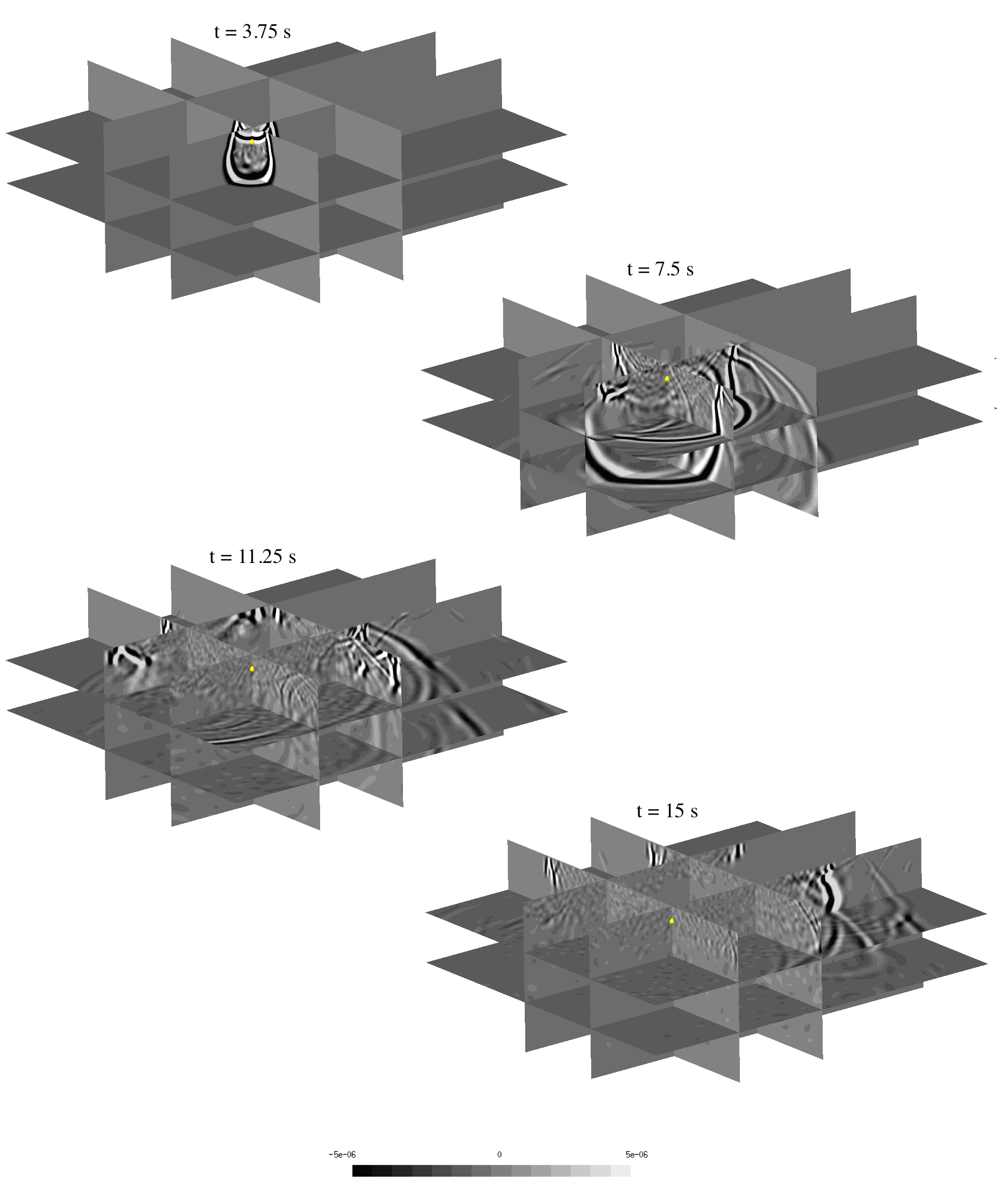}
\end{center}
\vspace{-0.4cm}
\caption{Snapshots of the pressure wavefield at different instants for the realistic benchmark.
  The free-surface boundary condition $(p=0)$ is used for the upper face of the domain,
  while the HABC with $N=3$ is used for the other faces.
  The location of the source point is represented with a yellow bullet on all the figures.}
\label{fig:SEAG:solution}
\end{figure}

We have performed simulations using the basic ABC and HABCs with $N=3$ and $N=6$.
Figure \ref{fig:SEAG:solution} shows snapshots of the solution at different instants for the HABC with $N=3$.
At $t=3.75\s$, we can see the primary wavefront, as well as the secondary wavefront generated after reflection on the free surface of the domain.
The source, represented with a yellow bullet, is in the upper part of the domain, which corresponds to an ocean.
In the remainder of the simulation, both wavefronts are propagated in the ocean and the subsurface, and multiple reflections appear due to geological structures.
Waves travel significantly faster in the subsurface than in the ocean.
 
\begin{figure}[!tbp]
\centering
\bigskip
\begin{tabular}{c@{\:}c}
\begin{subfigure}[b]{70mm}
  \centering
  \caption{P-wave velocity}
  \includegraphics[width=70mm]{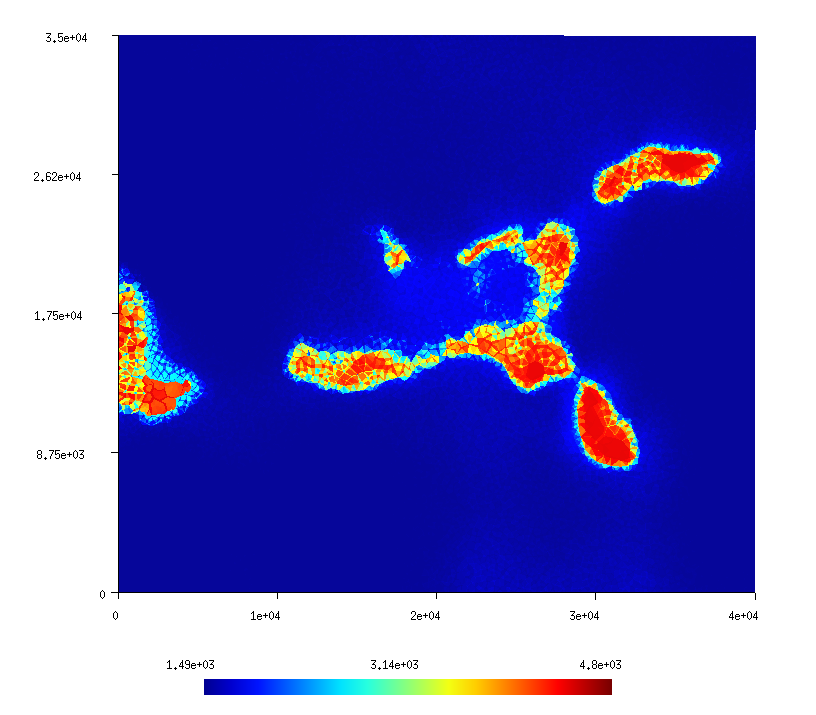}
  \label{fig:SEAM:rezuA:1}
\end{subfigure}
&
\begin{subfigure}[b]{70mm}
  \centering
  \caption{Density}
  \includegraphics[width=70mm]{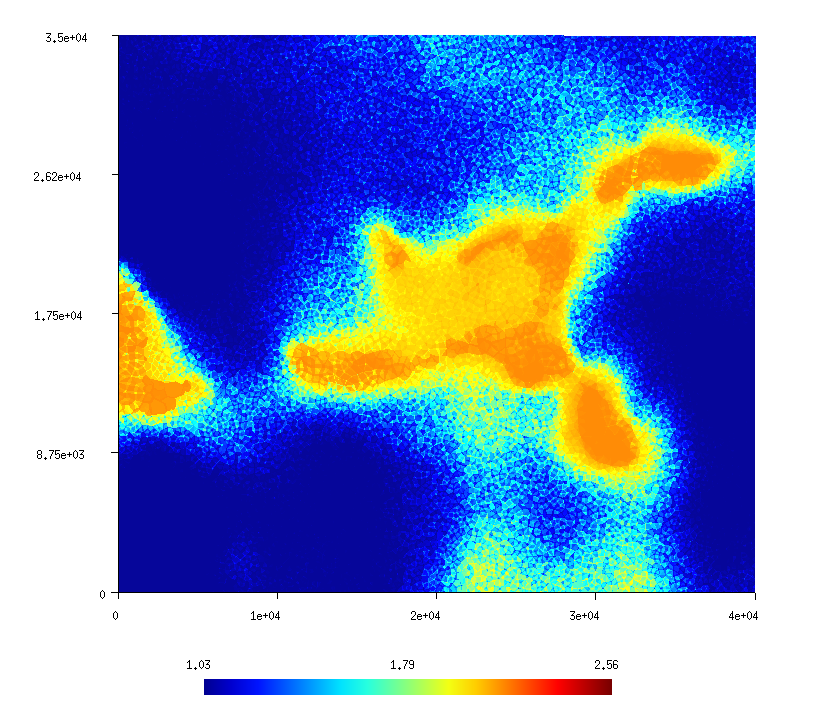}
  \label{fig:SEAM:rezuA:2}
\end{subfigure}
\vspace{0.25cm} \\
\begin{subfigure}[b]{70mm}
  \centering
  \caption{Pressure field for basic ABC}
  \includegraphics[width=70mm]{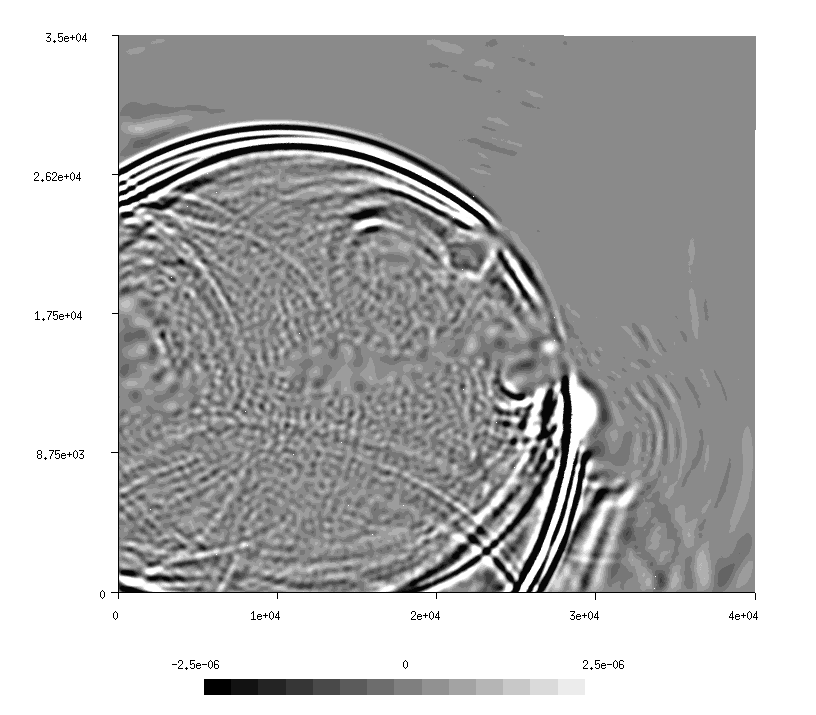}
  \label{fig:SEAM:rezuA:3}
\end{subfigure}
&
\begin{subfigure}[b]{70mm}
  \centering
  \caption{Pressure field for HABC with $N=3$}
  \includegraphics[width=70mm]{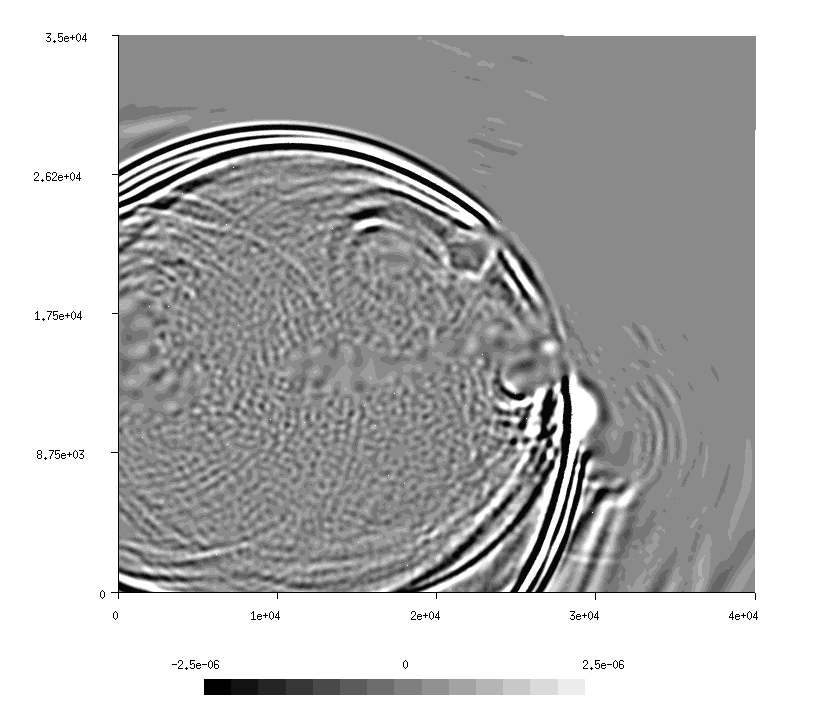}
  \label{fig:SEAM:rezuA:4}
\end{subfigure}
\end{tabular}
\vspace{0cm}
\caption{Snapshots of the medium properties and the pressure wavefield in an horizontal planar cut of the sea at depth $z=1.5\km$ and time $t=15\s$.
  The velocity and density models are represented in figures (a) and (b), respectively.
  The wavefield is computed using the basic ABC (c) and the HABC with $N=3$ (d).}
\label{fig:SEAM:rezuA}
\end{figure}

\begin{figure}[!tbp]
\centering
\bigskip
\begin{tabular}{c@{\:}c}
\begin{subfigure}[b]{70mm}
  \centering
  \caption{P-wave velocity}
  \includegraphics[width=70mm]{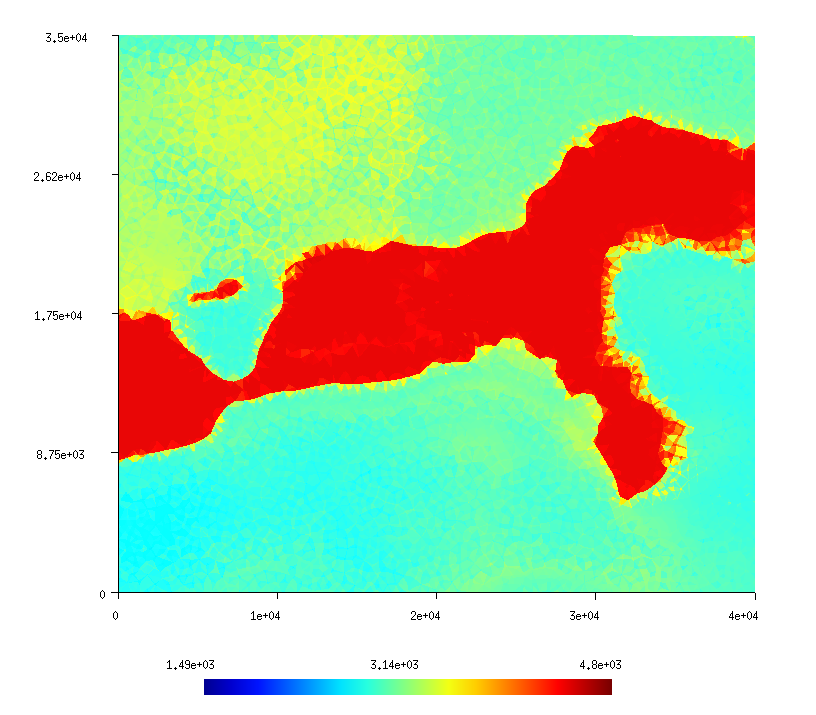}
  \label{fig:SEAM:rezuB:1}
\end{subfigure}
&
\begin{subfigure}[b]{70mm}
  \centering
  \caption{Density}
  \includegraphics[width=70mm]{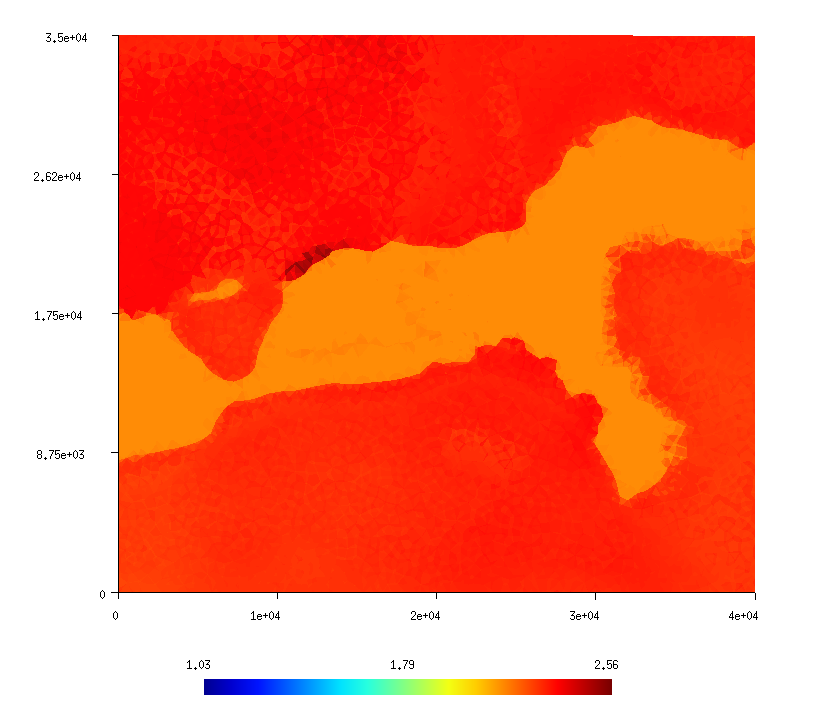}
  \label{fig:SEAM:rezuB:2}
\end{subfigure}
\vspace{0.25cm} \\
\begin{subfigure}[b]{70mm}
  \centering
  \caption{Pressure field for basic ABC}
  \includegraphics[width=70mm]{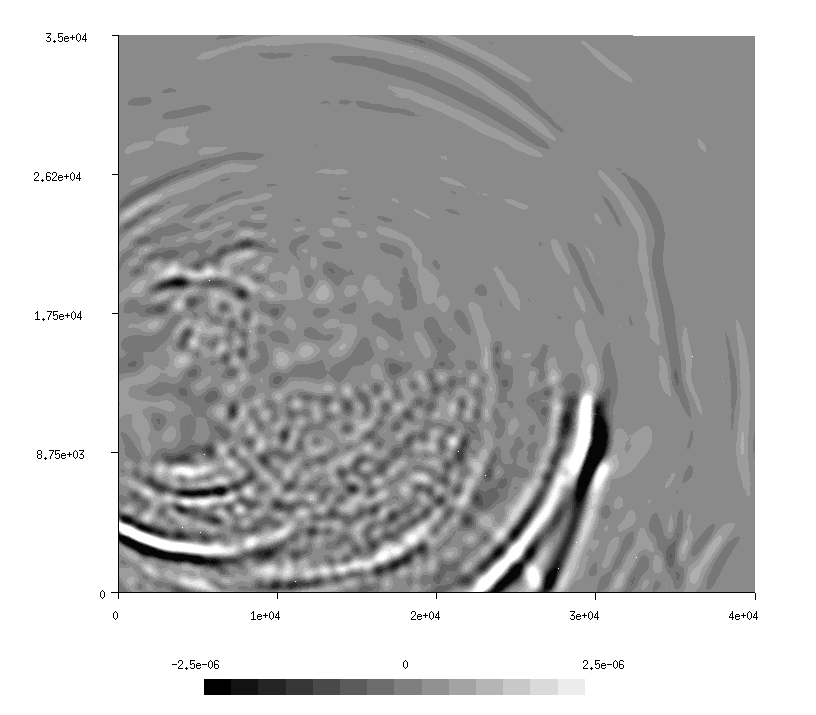}
  \label{fig:SEAM:rezuB:3}
\end{subfigure}
&
\begin{subfigure}[b]{70mm}
  \centering
  \caption{Pressure field for HABC with $N=3$}
  \includegraphics[width=70mm]{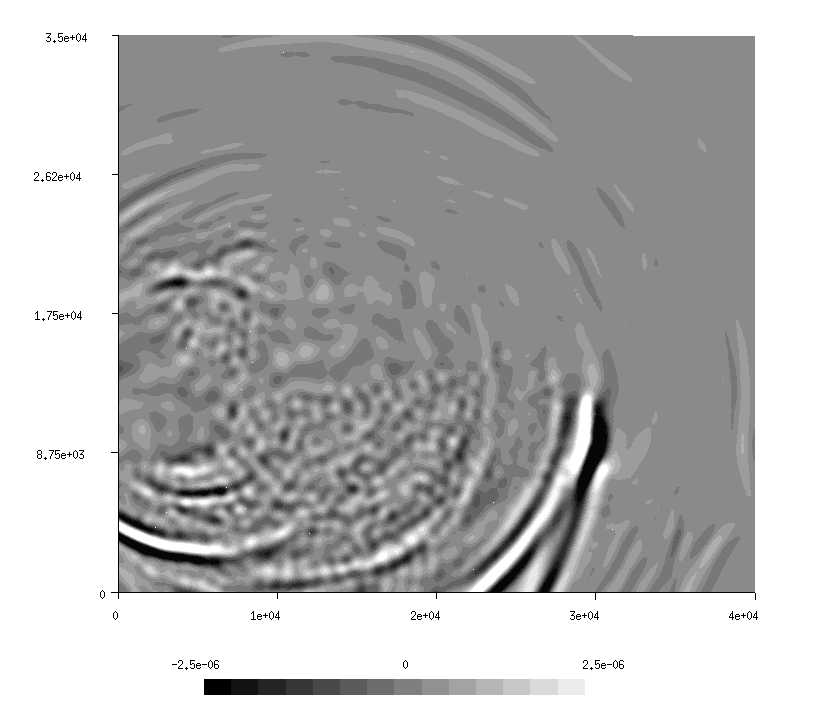}
  \label{fig:SEAM:rezuB:4}
\end{subfigure}
\end{tabular}
\vspace{0cm}
\caption{Snapshots of the medium properties and the pressure wavefield in an horizontal planar cut of the ground at depth $z=5\km$ and time $t=11.25\s$.
  The velocity and density models are represented in figures (a) and (b), respectively.
  The wavefield is computed using the basic ABC (c) and the HABC with $N=3$ (d).}
\label{fig:SEAM:rezuB}
\end{figure}

In order to compare the basic ABC with the HABCs, horizontal planar cuts of the pressure wavefield are shown in figures \ref{fig:SEAM:rezuA} and \ref{fig:SEAM:rezuB} at depths $z=1.5\km$ and $z=5\km$.
HABC results are shown only for $N=3$, because the images obtained with $N=6$ are visually nearly identical.

The horizontal cuts shown on figure \ref{fig:SEAM:rezuA} are taken at the ocean level where the source is placed, at the end of the simulation.
On the left border of the pictures, the medium is heterogeneous because the salt body touches the border (figures \ref{fig:SEAM:rezuA:1} and \ref{fig:SEAM:rezuA:2}).
There is therefore a large change of the medium properties that must be represented by the boundary treatment.
Comparing the results obtained with both boundary treatments, we can clearly observe reflections close to the left and lower sides of figure \ref{fig:SEAM:rezuA:3} (basic ABC) that are not on figure \ref{fig:SEAM:rezuA:4} (HABC).
The larger reflections correspond to oblique waves, while the wavefront corresponding to a normal incidence is not reflected by the basic ABC.
Note that the solution obtained with the HABC does not exhibit any incoherent behavior on the left side of the picture, where the medium is heterogeneous.

Figure \ref{fig:SEAM:rezuB} shows horizontal cuts.
These cuts are taken earlier in the simulation in order to observe eventual reflections of the primary wavefront at the boundary.
Again, we observe reflections when the ABC is used, and not with the HABC.


\subsubsection*{Computational performance}

The total runtime of the realistic simulation using the basic ABC is approximately 8h45 on a Nvidia K40 GPU in single precision.
Using the HABCs increases the runtime by 10 and 22 minutes for $N=3$ and $N=6$, respectively.
As shown on table \ref{table:SEAM:perf}, the runtime rises nearly proportionally to the total number of unknowns.

We have optimized the number of elements per thread block $K_\text{blk}$ for each kernel and each boundary treatment (see section \ref{sec:implementation}).
The optimum values are given in table \ref{table:SEAM:optiKblk}.
Optimizing these parameters $K_\text{blk}$ has a significant impact on the efficiency.
As shown on table \ref{table:SEAM:rezuRuntime}, the speedup achieved after optimization of $K_\text{blk}$ kernel by kernel is between 1.25 and 3.95.
The largest speedups are obtained with the 2D kernels and the 3D volume kernel.
The speedup of the complete implementation after optimization is approximately 1.85.

\begin{table}[!tb]
\begin{center}
\small
\begin{tabular}{|l|r|r|r|} \hline 
                            & Basic ABC & HABC \footnotesize$(N=3)$ & HABC \footnotesize$(N=6)$ \\ \hline
  Total number of unknowns  & 94,399,120 & 96,573,304 \small(+2.3\%) & 98,795,296 \small(+4.6\%) \\
  Total runtime             & 8h44 50' & 8h54 45' \small(+1.9\%) & 9h07 43' \small(+4.4\%) \\ \hline
\end{tabular}
\vspace{-0.25cm}
\end{center}
\caption{Performance statistics for the realistic benchmark when using a basic ABC or a HABC as boundary treatment.}
\label{table:SEAM:perf}
\end{table}

\begin{table}[!tb]
\begin{center}
\small
\begin{tabular}{|c|c|c|c|} \hline 
  Kernel & Basic ABC & HABC \footnotesize$(N=3)$ & HABC \footnotesize$(N=6)$ \\ \hline
  1D volume  & & 3 & 3 \\
  2D volume  & & 16 & 13 \\
  3D volume  & 8 & 8 & 8 \\
  1D surface & & 3 & 3 \\
  2D surface & & 8 & 5 \\
  3D surface & 2 & 2 & 2 \\
  1D update  & & 8 & 6 \\
  2D update  & & 10 & 8 \\
  3D update  & 3 & 3 & 3 \\ \hline
\end{tabular}
\vspace{-0.25cm}
\end{center}
\caption{Optimum value of the thread block size $K_\text{blk}$ for each kernel and each variant of the realistic benchmark.}
\label{table:SEAM:optiKblk}
\end{table}

\begin{table}[!tb]
\begin{center}
{\small
\begin{tabular}{|c|r|r|r|r|r|r|r|r|r|} \hline 
  Kernel &
  \multicolumn{3}{c|}{Basic ABC} &
  \multicolumn{3}{c|}{HABC \footnotesize$(N=3)$} &
  \multicolumn{3}{c|}{HABC \footnotesize$(N=6)$} \\ \hline
             & \footnotesize $K_\text{blk}=1$
             & \footnotesize Opti $K_\text{blk}$
             & \footnotesize SpUp
             & \footnotesize $K_\text{blk}=1$
             & \footnotesize Opti $K_\text{blk}$
             & \footnotesize SpUp
             & \footnotesize $K_\text{blk}=1$
             & \footnotesize Opti $K_\text{blk}$
             & \footnotesize SpUp \\ \hline
  1D volume  &            &          &      & $12.9\mus$ & $9.52\mus$ & 1.36 & $27.4\mus$ & $19.9\mus$ & 1.38 \\
  2D volume  &            &          &      &  $533\mus$ &  $135\mus$ & 3.95 &  $783\mus$ &  $305\mus$ & 2.57 \\
  3D volume  & $33.2\ms$ & $13.1\ms$ & 2.53 &  $33.5\ms$ &  $13.1\ms$ & 2.56 &  $33.4\ms$ &  $13.1\ms$ & 2.55 \\
  1D surface &            &          &      & $24.6\mus$ & $19.3\mus$ & 1.27 & $68.5\mus$ & $55.0\mus$ & 1.25 \\
  2D surface &            &          &      &  $870\mus$ &  $309\mus$ & 2.82 &  $1.42\ms$ &  $721\mus$ & 1.97 \\
  3D surface & $43.5\ms$ & $24.8\ms$ & 1.75 &  $44.0\ms$ &  $25.1\ms$ & 1.75 &  $44.0\ms$ &  $25.1\ms$ & 1.75 \\
  0D update  &            &          &      & $12.7\mus$ & $12.7\mus$ &      & $22.7\mus$ & $22.7\mus$ &      \\
  1D update  &            &          &      & $29.0\mus$ & $19.9\mus$ & 1.46 & $54.4\mus$ & $37.3\mus$ & 1.46 \\
  2D update  &            &          &      &  $958\mus$ &  $526\mus$ & 1.82 &  $1.53\ms$ &  $875\mus$ & 1.75 \\
  3D update  & $19.1\ms$ & $14.2\ms$ & 1.35 &  $19.2\ms$ &  $14.2\ms$ & 1.35 &  $19.2\ms$ &  $14.2\ms$ & 1.35 \\ \hline
\end{tabular}
}
\end{center}
\vspace{-0.25cm}
\caption{Average runtime of each kernel per call with and without optimized $K_\text{blk}$, and speedup (SpUp), for each variant of the realistic benchmark.
The 0D update kernel has no tuning parameter.}
\label{table:SEAM:rezuRuntime}
\end{table}


\newpage
\section{Conclusion}
\label{sec:conclusion}

A comprehensive computational procedure is proposed for the high-performance simulation of transient waves in 3D unbounded domains.
This approach combines a GPU-accelerated nodal discontinuous Galerkin finite element with local high-order absorbing boundary conditions (HABCs) and compatibility conditions for the edges and the corners of truncated cuboidal domains.
Since this approach can be naturally coupled with multi-rate time stepping schemes, discontinuous Galerkin schemes for hybrid meshes and computational strategies for computations on GPU clusters, our strategy has potential for large-scale realistic simulations that are both accurate and efficient on modern parallel architectures.

The considered HABC are variants of the classical Pad\'e-like approximate boundary conditions introduced by Engquist and Majda \cite{engquist1977absorbing} four decades ago.
With these conditions, outgoing traveling waves are simulated with an arbitrarily high accuracy, but adequate treatments must be designed to deal with the edges and the corners of cuboidal domains.
By choosing a specific representation for the HABC, we have derived novel edge/corner compatibility conditions that are rather naturally coupled with discontinuous Galerkin methods.
This is in contrast to classical HABC representations, which lead to inconsistency when applied to discontinuous Galerkin schemes.
Boundary formulations have been derived for the wave equation and the pressure-velocity system defined on the infinite space $\mathbb{R}^3$, assuming a homogeneous medium in the exterior domain $\mathbb{R}^3\backslash\Omega$.
Homogeneous boundary conditions are straightforwardly incorporated in these formulations, and numerical results suggest that they can be used with heterogeneous media.

The computational procedure relies on a multidimensional solver with partial differential equations to solve in the volume, on the faces and the edges of the computational domain.
When using a formulation based on the pressure-velocity system, the procedure can be performed purely explicitly using classical explicit time-stepping schemes and existing spatial schemes for the 1D, 2D and 3D versions of the pressure-velocity system.
We have proposed a GPU computational implementation based on a Runge-Kutta time-stepping scheme and a nodal discontinuous Galerkin method.
For each of the 1D, 2D and 3D parts of the solver, we have used optimization strategies which leverage the discrete structure of nodal discontinuous Galerkin schemes and speed the final implementation.
Numerical and computational results confirm the applicability and the efficiency of the approach.

As natural extensions of this work, we plan to derive similar HABCs with edge/corner compatibility conditions for other wave equations.
These formulations can be based on the Pad\'e-like approximate boundary conditions already proposed in the literature for electromagnetic \cite{el2014approximate} and elastic waves \cite{chaillat2015approximate} in the frequency domain.
We also plan to investigate variations of the HABCs that we have used with the aim of addressing long-time instabilities, and to study connections with existing long-time stable absorbing boundary conditions (see \eg \cite{hagstrom2009complete,hagstrom2010radiation,baffet2012long}).

\section*{Acknowledgements}

This work was funded by a grant from TOTAL E\&P Research and Technology USA.
The authors thank TOTAL for permission to publish.
Axel Modave was partially supported by an excellence grant from Wallonie-Bruxelles International (WBI), and was a Postdoctoral Researcher on leave with the F.R.S-FNRS.
The first author thanks Thomas Hagstrom for helpful and informative discussions.


\small
\setlength{\bibsep}{0pt plus 0ex}
\bibliographystyle{abbrvnat}
\bibliography{myrefs}

\begin{thebibliography}{72}
\providecommand{\natexlab}[1]{#1}
\providecommand{\url}[1]{\texttt{#1}}
\expandafter\ifx\csname urlstyle\endcsname\relax
  \providecommand{\doi}[1]{doi: #1}\else
  \providecommand{\doi}{doi: \begingroup \urlstyle{rm}\Url}\fi

\bibitem[Antoine et~al.(2006)Antoine, Darbas, and Lu]{antoine2006improved}
X.~Antoine, M.~Darbas, and Y.~Y. Lu.
\newblock An improved surface radiation condition for high-frequency acoustic
  scattering problems.
\newblock \emph{Computer Methods in Applied Mechanics and Engineering},
  195\penalty0 (33):\penalty0 4060--4074, 2006.

\bibitem[Appel{\"o} et~al.(2006)Appel{\"o}, Hagstrom, and
  Kreiss]{appelo2006perfectly}
D.~Appel{\"o}, T.~Hagstrom, and G.~Kreiss.
\newblock Perfectly matched layers for hyperbolic systems: general formulation,
  well-posedness, and stability.
\newblock \emph{SIAM Journal on Applied Mathematics}, 67\penalty0 (1):\penalty0
  1--23, 2006.

\bibitem[Asvadurov et~al.(2003)Asvadurov, Druskin, Guddati, and
  Knizhnerman]{asvadurov2003optimal}
S.~Asvadurov, V.~Druskin, M.~N. Guddati, and L.~Knizhnerman.
\newblock On optimal finite-difference approximation of pml.
\newblock \emph{SIAM Journal on Numerical Analysis}, 41\penalty0 (1):\penalty0
  287--305, 2003.

\bibitem[Baffet et~al.(2012)Baffet, Bielak, Givoli, Hagstrom, and
  Rabinovich]{baffet2012long}
D.~Baffet, J.~Bielak, D.~Givoli, T.~Hagstrom, and D.~Rabinovich.
\newblock Long-time stable high-order absorbing boundary conditions for
  elastodynamics.
\newblock \emph{Computer Methods in Applied Mechanics and Engineering},
  241:\penalty0 20--37, 2012.

\bibitem[Baffet et~al.(2014)Baffet, Hagstrom, and Givoli]{baffet2014double}
D.~Baffet, T.~Hagstrom, and D.~Givoli.
\newblock Double absorbing boundary formulations for acoustics and
  elastodynamics.
\newblock \emph{SIAM Journal on Scientific Computing}, 36\penalty0
  (3):\penalty0 A1277--A1312, 2014.

\bibitem[Bamberger et~al.(1988)Bamberger, Engquist, Halpern, and
  Joly]{bamberger1988higher}
A.~Bamberger, B.~Engquist, L.~Halpern, and P.~Joly.
\newblock Higher order paraxial wave equation approximations in heterogeneous
  media.
\newblock \emph{SIAM Journal on Applied Mathematics}, 48\penalty0 (1):\penalty0
  129--154, 1988.

\bibitem[B{\'e}cache et~al.(2010)B{\'e}cache, Givoli, and
  Hagstrom]{becache2010high}
E.~B{\'e}cache, D.~Givoli, and T.~Hagstrom.
\newblock High-order absorbing boundary conditions for anisotropic and
  convective wave equations.
\newblock \emph{Journal of Computational Physics}, 229\penalty0 (4):\penalty0
  1099--1129, 2010.

\bibitem[B\'erenger(1994)]{berenger1994perfectly}
J.-P. B\'erenger.
\newblock A perfectly matched layer for the absorption of electromagnetic
  waves.
\newblock \emph{Journal of Computational Physics}, 114\penalty0 (2):\penalty0
  185--200, 1994.

\bibitem[Berm{\'u}dez et~al.(2007)Berm{\'u}dez, Hervella-Nieto, Prieto, and
  Rodriguez]{bermudez2007optimal}
A.~Berm{\'u}dez, L.~Hervella-Nieto, A.~Prieto, and R.~Rodriguez.
\newblock An optimal perfectly matched layer with unbounded absorbing function
  for time-harmonic acoustic scattering problems.
\newblock \emph{Journal of Computational Physics}, 223\penalty0 (2):\penalty0
  469--488, 2007.

\bibitem[Carpenter and Kennedy(1994)]{carpenter1994fourth}
M.~H. Carpenter and C.~A. Kennedy.
\newblock {Fourth-order 2N-storage Runge-Kutta schemes}.
\newblock Technical Report NASA-TM-109112, NASA Langley Research Center, 1994.

\bibitem[Chaillat et~al.(2015)Chaillat, Darbas, and
  Le~Lou{\"e}r]{chaillat2015approximate}
S.~Chaillat, M.~Darbas, and F.~Le~Lou{\"e}r.
\newblock Approximate local {D}irichlet-to-{N}eumann map for three-dimensional
  time-harmonic elastic waves.
\newblock \emph{Computer Methods in Applied Mechanics and Engineering},
  297:\penalty0 62--83, 2015.

\bibitem[Chan and Warburton(2015)]{chan2015gpu}
J.~Chan and T.~Warburton.
\newblock {GPU}-accelerated {B}ernstein-{B}ezier discontinuous {G}alerkin
  methods for wave problems.
\newblock \emph{arXiv preprint arXiv:1512.06025}, 2015.

\bibitem[Chan et~al.(2016{\natexlab{a}})Chan, Wang, Hewett, and
  Warburton]{chan2016reduced}
J.~Chan, Z.~Wang, R.~J. Hewett, and T.~Warburton.
\newblock Reduced storage nodal discontinuous {G}alerkin methods on
  semi-structured prismatic meshes.
\newblock \emph{arXiv preprint arXiv:1607.03399}, 2016{\natexlab{a}}.

\bibitem[Chan et~al.(2016{\natexlab{b}})Chan, Wang, Modave, Remacle, and
  Warburton]{chan2016gpu}
J.~Chan, Z.~Wang, A.~Modave, J.-F. Remacle, and T.~Warburton.
\newblock {GPU}-accelerated discontinuous {G}alerkin methods on hybrid meshes.
\newblock \emph{Journal of Computational Physics}, 318:\penalty0 142--168,
  2016{\natexlab{b}}.

\bibitem[Cockburn et~al.(2000)Cockburn, Karniadakis, and
  Shu]{cockburn2000development}
B.~Cockburn, G.~E. Karniadakis, and C.-W. Shu.
\newblock The development of discontinuous {G}alerkin methods.
\newblock In \emph{Discontinuous Galerkin Methods}, pages 3--50. Springer,
  2000.

\bibitem[Collino(1993{\natexlab{a}})]{collino1993conditions}
F.~Collino.
\newblock Conditions absorbantes d'ordre {\'e}lev{\'e} pour les {\'e}quations
  de maxwell dans des domaines rectangulaires.
\newblock Technical Report 2932, INRIA, 1993{\natexlab{a}}.

\bibitem[Collino(1993{\natexlab{b}})]{collino1993high}
F.~Collino.
\newblock High order absorbing boundary conditions for wave propagation models.
  {S}traight line boundary and corner cases.
\newblock In \emph{Second International Conference on Mathematical and
  Numerical Aspects of Wave Propagation (Newark, DE, 1993)}, pages 161--171,
  1993{\natexlab{b}}.

\bibitem[Collino and Monk(1998)]{collino1998optimizing}
F.~Collino and P.~B. Monk.
\newblock Optimizing the perfectly matched layer.
\newblock \emph{Computer Methods in Applied Mechanics and Engineering},
  164\penalty0 (1):\penalty0 157--171, 1998.

\bibitem[El~Bouajaji et~al.(2014)El~Bouajaji, Antoine, and
  Geuzaine]{el2014approximate}
M.~El~Bouajaji, X.~Antoine, and C.~Geuzaine.
\newblock Approximate local magnetic-to-electric surface operators for
  time-harmonic maxwell's equations.
\newblock \emph{Journal of Computational Physics}, 279:\penalty0 241--260,
  2014.

\bibitem[Engquist and Majda(1977)]{engquist1977absorbing}
B.~Engquist and A.~Majda.
\newblock Absorbing boundary conditions for numerical simulation of waves.
\newblock \emph{Proceedings of the National Academy of Sciences}, 74\penalty0
  (5):\penalty0 1765--1766, 1977.

\bibitem[Engquist and Majda(1979)]{engquist1979radiation}
B.~Engquist and A.~Majda.
\newblock Radiation boundary conditions for acoustic and elastic wave
  calculations.
\newblock \emph{Communications on pure and applied mathematics}, 32\penalty0
  (3):\penalty0 313--357, 1979.

\bibitem[Fehler and Keliher(2011)]{fehler2011seam}
M.~Fehler and P.~J. Keliher.
\newblock \emph{{SEAM} {P}hase 1: Challenges of Subsalt Imaging in Tertiary
  Basins, with Emphasis on Deepwater {G}ulf of {M}exico}.
\newblock Society of Exploration Geophysicists Tulsa, 2011.

\bibitem[Fuhry et~al.(2014)Fuhry, Giuliani, and
  Krivodonova]{fuhry2014discontinuous}
M.~Fuhry, A.~Giuliani, and L.~Krivodonova.
\newblock Discontinuous {G}alerkin methods on graphics processing units for
  nonlinear hyperbolic conservation laws.
\newblock \emph{International Journal for Numerical Methods in Fluids},
  76\penalty0 (12):\penalty0 982--1003, 2014.

\bibitem[Gandham et~al.(2015)Gandham, Medina, and Warburton]{gandham2015gpu}
R.~Gandham, D.~Medina, and T.~Warburton.
\newblock {GPU} accelerated discontinuous {G}alerkin methods for shallow water
  equations.
\newblock \emph{Communications in Computational Physics}, 18\penalty0
  (1):\penalty0 37--64, 2015.

\bibitem[Gedney(1996)]{gedney1996anisotropic}
S.~D. Gedney.
\newblock An anisotropic perfectly matched layer-absorbing medium for the
  truncation of {FDTD} lattices.
\newblock \emph{IEEE transactions on Antennas and Propagation}, 44\penalty0
  (12):\penalty0 1630--1639, 1996.

\bibitem[Givoli(2001)]{givoli2001high}
D.~Givoli.
\newblock High-order nonreflecting boundary conditions without high-order
  derivatives.
\newblock \emph{Journal of Computational Physics}, 170\penalty0 (2):\penalty0
  849--870, 2001.

\bibitem[Givoli(2004)]{givoli2004high}
D.~Givoli.
\newblock High-order local non-reflecting boundary conditions: a review.
\newblock \emph{Wave Motion}, 39\penalty0 (4):\penalty0 319--326, 2004.

\bibitem[Givoli and Neta(2003)]{givoli2003high}
D.~Givoli and B.~Neta.
\newblock High-order non-reflecting boundary scheme for time-dependent waves.
\newblock \emph{Journal of Computational Physics}, 186\penalty0 (1):\penalty0
  24--46, 2003.

\bibitem[Givoli et~al.(1997)Givoli, Patlashenko, and Keller]{givoli1997high}
D.~Givoli, I.~Patlashenko, and J.~B. Keller.
\newblock High-order boundary conditions and finite elements for infinite
  domains.
\newblock \emph{Computer Methods in Applied Mechanics and Engineering},
  143\penalty0 (1):\penalty0 13--39, 1997.

\bibitem[Godel et~al.(2010)Godel, Nunn, Warburton, and
  Clemens]{godel2010scalability}
N.~Godel, N.~Nunn, T.~Warburton, and M.~Clemens.
\newblock Scalability of higher-order discontinuous {G}alerkin {FEM}
  computations for solving electromagnetic wave propagation problems on {GPU}
  clusters.
\newblock \emph{IEEE Transactions on Magnetics}, 46\penalty0 (8):\penalty0
  3469--3472, 2010.

\bibitem[G\"odel et~al.(2010)G\"odel, Schomann, Warburton, and
  Clemens]{godel2010gpu}
N.~G\"odel, S.~Schomann, T.~Warburton, and M.~Clemens.
\newblock {GPU} accelerated {A}dams--{B}ashforth multirate discontinuous
  {G}alerkin {FEM} simulation of high-frequency electromagnetic fields.
\newblock \emph{IEEE Transactions on magnetics}, 46\penalty0 (8):\penalty0
  2735--2738, 2010.

\bibitem[Guan-Quan(1985)]{guan1985high}
Z.~Guan-Quan.
\newblock High-order approximation of one way wave equations.
\newblock \emph{J. Comput. Math}, 3:\penalty0 90--97, 1985.

\bibitem[Guddati and Tassoulas(2000)]{guddati2000continued}
M.~N. Guddati and J.~L. Tassoulas.
\newblock Continued-fraction absorbing boundary conditions for the wave
  equation.
\newblock \emph{Journal of Computational Acoustics}, 8\penalty0 (01):\penalty0
  139--156, 2000.

\bibitem[Ha-Duong and Joly(1994)]{ha1994stability}
T.~Ha-Duong and P.~Joly.
\newblock On the stability analysis of boundary conditions for the wave
  equation by energy methods. {I}. {T}he homogeneous case.
\newblock \emph{Mathematics of Computation}, 62\penalty0 (206):\penalty0
  539--563, 1994.

\bibitem[Hagstrom(1999)]{hagstrom1999radiation}
T.~Hagstrom.
\newblock Radiation boundary conditions for the numerical simulation of waves.
\newblock \emph{Acta numerica}, 8:\penalty0 47--106, 1999.

\bibitem[Hagstrom and Warburton(2004)]{hagstrom2004new}
T.~Hagstrom and T.~Warburton.
\newblock A new auxiliary variable formulation of high-order local radiation
  boundary conditions: corner compatibility conditions and extensions to
  first-order systems.
\newblock \emph{Wave Motion}, 39\penalty0 (4):\penalty0 327--338, 2004.

\bibitem[Hagstrom and Warburton(2009)]{hagstrom2009complete}
T.~Hagstrom and T.~Warburton.
\newblock Complete radiation boundary conditions: minimizing the long time
  error growth of local methods.
\newblock \emph{SIAM Journal on Numerical Analysis}, 47\penalty0 (5):\penalty0
  3678--3704, 2009.

\bibitem[Hagstrom et~al.(2007)Hagstrom, De~Castro, Givoli, and
  Tzemach]{hagstrom2007local}
T.~Hagstrom, M.~L. De~Castro, D.~Givoli, and D.~Tzemach.
\newblock Local high-order absorbing boundary conditions for time-dependent
  waves in guides.
\newblock \emph{Journal of Computational Acoustics}, 15\penalty0 (01):\penalty0
  1--22, 2007.

\bibitem[Hagstrom et~al.(2010)Hagstrom, Warburton, and
  Givoli]{hagstrom2010radiation}
T.~Hagstrom, T.~Warburton, and D.~Givoli.
\newblock Radiation boundary conditions for time-dependent waves based on
  complete plane wave expansions.
\newblock \emph{Journal of Computational and Applied Mathematics}, 234\penalty0
  (6):\penalty0 1988--1995, 2010.

\bibitem[Hagstrom et~al.(2014)Hagstrom, Givoli, Rabinovich, and
  Bielak]{hagstrom2014double}
T.~Hagstrom, D.~Givoli, D.~Rabinovich, and J.~Bielak.
\newblock The double absorbing boundary method.
\newblock \emph{Journal of Computational Physics}, 259:\penalty0 220--241,
  2014.

\bibitem[Halpern and Trefethen(1988)]{halpern1988wide}
L.~Halpern and L.~N. Trefethen.
\newblock Wide-angle one-way wave equations.
\newblock \emph{The Journal of the Acoustical Society of America}, 84\penalty0
  (4):\penalty0 1397--1404, 1988.

\bibitem[Hesthaven and Warburton(2007)]{hesthaven2007nodal}
J.~S. Hesthaven and T.~Warburton.
\newblock \emph{Nodal discontinuous Galerkin methods: algorithms, analysis, and
  applications}.
\newblock Springer Science \& Business Media, 2007.

\bibitem[Higdon(1986)]{higdon1986absorbing}
R.~L. Higdon.
\newblock Absorbing boundary conditions for difference approximations to the
  multidimensional wave equation.
\newblock \emph{Mathematics of computation}, 47\penalty0 (176):\penalty0
  437--459, 1986.

\bibitem[Hu(2008)]{hu2008development}
F.~Q. Hu.
\newblock Development of {PML} absorbing boundary conditions for computational
  aeroacoustics: {A} progress review.
\newblock \emph{Computers \& Fluids}, 37\penalty0 (4):\penalty0 336--348, 2008.

\bibitem[Ingerman et~al.(2000)Ingerman, Druskin, and
  Knizhnerman]{ingerman2000optimal}
D.~Ingerman, V.~Druskin, and L.~Knizhnerman.
\newblock Optimal finite difference grids and rational approximations of the
  square root i. elliptic problems.
\newblock \emph{Communications on Pure and Applied Mathematics}, 53\penalty0
  (8):\penalty0 1039--1066, 2000.

\bibitem[Kechroud et~al.(2005)Kechroud, Antoine, and
  Soulaimani]{kechroud2005numerical}
R.~Kechroud, X.~Antoine, and A.~Soulaimani.
\newblock Numerical accuracy of a pad{\'e}-type non-reflecting boundary
  condition for the finite element solution of acoustic scattering problems at
  high-frequency.
\newblock \emph{International Journal for Numerical Methods in Engineering},
  64\penalty0 (10):\penalty0 1275--1302, 2005.

\bibitem[Kl\"ockner et~al.(2009)Kl\"ockner, Warburton, Bridge, and
  Hesthaven]{klockner2009nodal}
A.~Kl\"ockner, T.~Warburton, J.~Bridge, and J.~S. Hesthaven.
\newblock Nodal discontinuous {G}alerkin methods on graphics processors.
\newblock \emph{Journal of Computational Physics}, 228\penalty0 (21):\penalty0
  7863--7882, 2009.

\bibitem[Komatitsch and Tromp(2003)]{komatitsch2003perfectly}
D.~Komatitsch and J.~Tromp.
\newblock A perfectly matched layer absorbing boundary condition for the
  second-order seismic wave equation.
\newblock \emph{Geophysical Journal International}, 154\penalty0 (1):\penalty0
  146--153, 2003.

\bibitem[LaGrone and Hagstrom(2016)]{lagrone2016double}
J.~LaGrone and T.~Hagstrom.
\newblock Double absorbing boundaries for finite-difference time-domain
  electromagnetics.
\newblock \emph{Journal of Computational Physics}, 326:\penalty0 650--665,
  2016.

\bibitem[L{\'e}ger et~al.(2014)L{\'e}ger, Viquerat, Durochat, Scheid, and
  Lanteri]{leger2014parallel}
R.~L{\'e}ger, J.~Viquerat, C.~Durochat, C.~Scheid, and S.~Lanteri.
\newblock A parallel non-conforming multi-element {DGTD} method for the
  simulation of electromagnetic wave interaction with metallic nanoparticles.
\newblock \emph{Journal of Computational and Applied Mathematics},
  270:\penalty0 330--342, 2014.

\bibitem[LeVeque(2002)]{leveque2002finite}
R.~J. LeVeque.
\newblock \emph{Finite volume methods for hyperbolic problems}, volume~31.
\newblock Cambridge university press, 2002.

\bibitem[Li and Hesthaven(2014)]{li2014analysis}
J.~Li and J.~S. Hesthaven.
\newblock Analysis and application of the nodal discontinuous {G}alerkin method
  for wave propagation in metamaterials.
\newblock \emph{Journal of Computational Physics}, 258:\penalty0 915--930,
  2014.

\bibitem[Lu et~al.(2004)Lu, Zhang, and Cai]{lu2004discontinuous}
T.~Lu, P.~Zhang, and W.~Cai.
\newblock Discontinuous galerkin methods for dispersive and lossy maxwell's
  equations and pml boundary conditions.
\newblock \emph{Journal of Computational Physics}, 200\penalty0 (2):\penalty0
  549--580, 2004.

\bibitem[Lu(1998)]{lu1998complex}
Y.~Y. Lu.
\newblock A complex coefficient rational approximation of $\sqrt{1+x}$.
\newblock \emph{Applied numerical mathematics}, 27\penalty0 (2):\penalty0
  141--154, 1998.

\bibitem[Medina et~al.(2014)Medina, St.{-}Cyr, and Warburton]{medina2014occa}
D.~S. Medina, A.~St.{-}Cyr, and T.~Warburton.
\newblock {OCCA:} {A} unified approach to multi-threading languages.
\newblock 2014.
\newblock http://arxiv.org/abs/1403.0968.

\bibitem[Mercerat and Glinsky(2015)]{mercerat2015nodal}
E.~D. Mercerat and N.~Glinsky.
\newblock A nodal high-order discontinuous galerkin method for elastic wave
  propagation in arbitrary heterogeneous media.
\newblock \emph{Geophysical Journal International}, 201\penalty0 (2):\penalty0
  1101--1118, 2015.

\bibitem[Milinazzo et~al.(1997)Milinazzo, Zala, and
  Brooke]{milinazzo1997rational}
F.~A. Milinazzo, C.~A. Zala, and G.~H. Brooke.
\newblock Rational square-root approximations for parabolic equation
  algorithms.
\newblock \emph{The Journal of the Acoustical Society of America}, 101\penalty0
  (2):\penalty0 760--766, 1997.

\bibitem[Modave et~al.(2014)Modave, Delhez, and Geuzaine]{modave2014optimizing}
A.~Modave, E.~Delhez, and C.~Geuzaine.
\newblock Optimizing perfectly matched layers in discrete contexts.
\newblock \emph{International Journal for Numerical Methods in Engineering},
  99\penalty0 (6):\penalty0 410--437, 2014.

\bibitem[Modave et~al.(2015)Modave, St-Cyr, Mulder, and
  Warburton]{modave2015nodal}
A.~Modave, A.~St-Cyr, W.~A. Mulder, and T.~Warburton.
\newblock A nodal discontinuous {G}alerkin method for reverse-time migration on
  {GPU} clusters.
\newblock \emph{Geophysical Journal International}, 203\penalty0 (2):\penalty0
  1419--1435, 2015.

\bibitem[Modave et~al.(2016{\natexlab{a}})Modave, Lambrechts, and
  Geuzaine]{modave2016perfectly}
A.~Modave, J.~Lambrechts, and C.~Geuzaine.
\newblock Perfectly matched layers for convex truncated domains with
  discontinuous {G}alerkin time domain simulations, 2016{\natexlab{a}}.
\newblock Manuscript submitted for publication.

\bibitem[Modave et~al.(2016{\natexlab{b}})Modave, St-Cyr, and
  Warburton]{modave2016gpu}
A.~Modave, A.~St-Cyr, and T.~Warburton.
\newblock {GPU} performance analysis of a nodal discontinuous {G}alerkin method
  for acoustic and elastic models.
\newblock \emph{Computers \& Geosciences}, 91:\penalty0 64--76,
  2016{\natexlab{b}}.

\bibitem[Rabinovich et~al.(2011)Rabinovich, Givoli, Bielak, and
  Hagstrom]{rabinovich2011finite}
D.~Rabinovich, D.~Givoli, J.~Bielak, and T.~Hagstrom.
\newblock A finite element scheme with a high order absorbing boundary
  condition for elastodynamics.
\newblock \emph{Computer Methods in Applied Mechanics and Engineering},
  200\penalty0 (23):\penalty0 2048--2066, 2011.

\bibitem[Schmidt et~al.(2015)Schmidt, Diaz, and Heier]{schmidt2015non}
K.~Schmidt, J.~Diaz, and C.~Heier.
\newblock Non-conforming {G}alerkin finite element methods for local absorbing
  boundary conditions of higher order.
\newblock \emph{Computers \& Mathematics with Applications}, 70\penalty0
  (9):\penalty0 2252--2269, 2015.

\bibitem[Schmitt et~al.(2016)Schmitt, Scheid, Lanteri, Moreau, and
  Viquerat]{schmitt2016dgtd}
N.~Schmitt, C.~Scheid, S.~Lanteri, A.~Moreau, and J.~Viquerat.
\newblock A {DGTD} method for the numerical modeling of the interaction of
  light with nanometer scale metallic structures taking into account non-local
  dispersion effects.
\newblock \emph{Journal of Computational Physics}, 316:\penalty0 396--415,
  2016.

\bibitem[Seny et~al.(2014)Seny, Lambrechts, Toulorge, Legat, and
  Remacle]{seny2014efficient}
B.~Seny, J.~Lambrechts, T.~Toulorge, V.~Legat, and J.-F. Remacle.
\newblock An efficient parallel implementation of explicit multirate
  runge--kutta schemes for discontinuous galerkin computations.
\newblock \emph{Journal of Computational Physics}, 256:\penalty0 135--160,
  2014.

\bibitem[Toulorge and Desmet(2010)]{toulorge2010curved}
T.~Toulorge and W.~Desmet.
\newblock Curved boundary treatments for the discontinuous galerkin method
  applied to aeroacoustic propagation.
\newblock \emph{AIAA journal}, 48\penalty0 (2):\penalty0 479--489, 2010.

\bibitem[Warburton(2000)]{warburton2000application}
T.~Warburton.
\newblock Application of the discontinuous {G}alerkin method to {M}axwell’s
  equations using unstructured polymorphic hp-finite elements.
\newblock In \emph{Discontinuous Galerkin Methods}, pages 451--458. Springer,
  2000.

\bibitem[Warburton(2006)]{warburton2006explicit}
T.~Warburton.
\newblock An explicit construction of interpolation nodes on the simplex.
\newblock \emph{Journal of engineering mathematics}, 56\penalty0 (3):\penalty0
  247--262, 2006.

\bibitem[Warburton(2013)]{warburton2013low}
T.~Warburton.
\newblock A low-storage curvilinear discontinuous {G}alerkin method for wave
  problems.
\newblock \emph{SIAM Journal on Scientific Computing}, 35\penalty0
  (4):\penalty0 A1987--A2012, 2013.

\bibitem[Warburton and Hesthaven(2003)]{warburton2003constants}
T.~Warburton and J.~S. Hesthaven.
\newblock On the constants in hp-finite element trace inverse inequalities.
\newblock \emph{Computer Methods in Applied Mechanics and Engineering},
  192\penalty0 (25):\penalty0 2765--2773, 2003.

\bibitem[Wilcox et~al.(2010)Wilcox, Stadler, Burstedde, and
  Ghattas]{wilcox2010high}
L.~C. Wilcox, G.~Stadler, C.~Burstedde, and O.~Ghattas.
\newblock A high-order discontinuous {G}alerkin method for wave propagation
  through coupled elastic--acoustic media.
\newblock \emph{Journal of Computational Physics}, 229\penalty0 (24):\penalty0
  9373--9396, 2010.

\bibitem[Ye et~al.(2016)Ye, Maarten, Petrovitch, Pyrak-Nolte, and
  Wilcox]{ye2016discontinuous}
R.~Ye, V.~Maarten, C.~L. Petrovitch, L.~J. Pyrak-Nolte, and L.~C. Wilcox.
\newblock A discontinuous {G}alerkin method with a modified penalty flux for
  the propagation and scattering of acousto-elastic waves.
\newblock \emph{Geophysical Journal International}, 205\penalty0 (2):\penalty0
  1267--1289, 2016.

\end{thebibliography}

\end{document}